\documentclass[%
 aip,
 amsmath,amssymb,
 reprint,%
]{revtex4-1}

\usepackage{graphicx}
\usepackage{dcolumn}
\usepackage{bm}

\usepackage[utf8]{inputenc}
\usepackage[T1]{fontenc}
\usepackage{mathptmx}
\usepackage{etoolbox}
\usepackage[version=4]{mhchem}
\usepackage{float}
\usepackage{subfigure}
\usepackage{xcolor}
\usepackage{siunitx}
\usepackage{subcaption}
\definecolor{darkred}{rgb}{0.55,0,0}

\definecolor{chkgreen}{rgb}{0,0.5,0}
\newcommand{\chk}[1]{\textcolor{chkgreen}{#1}}
\usepackage[normalem]{ulem}

\makeatletter
\def\@email#1#2{%
 \endgroup
 \patchcmd{\titleblock@produce}
  {\frontmatter@RRAPformat}
  {\frontmatter@RRAPformat{\produce@RRAP{*#1\href{mailto:#2}{#2}}}\frontmatter@RRAPformat}
  {}{}
}%

\makeatother
\begin{document}

\preprint{AIP/123-QED}

\title[]{An Ab Initio Quantum-Chemistry-Based Three-Temperature Approach for Non-Equilibrium Hypersonic Flows: Application to Homogeneous 0-D Chemical Systems}
\author{Aakanksha Notey}
\altaffiliation[PhD student]{}
\affiliation{Center for Hypersonics and Entry Systems Studies (CHESS), \\ University of Illinois at Urbana-Champaign, Urbana, IL 61801, USA}

\author{Sung Min Jo}
\altaffiliation[Assistant Professor]{}
\affiliation{UCF Center of Excellence in Hypersonics and Space Propulsion (HYPERSPACE), \\ Department of Mechanical and Aerospace Engineering, \\ University of Central Florida, Orlando, FL 32816, USA}

\author{Marco Panesi}
\altaffiliation[Professor, Corresponding author]{}
\email{mpanesi@uci.edu}
\affiliation{Center for Hypersonics and Entry Systems Studies (CHESS), \\ University of Illinois at Urbana-Champaign, Urbana, IL 61801, USA}
\affiliation{Department of Mechanical and Aerospace Engineering, \\ University of California Irvine, Irvine, CA 92697, USA}

\date{\today}

\begin{abstract}
This work presents an ab initio quantum-chemistry-based one-bin, three-temperature (1-bin 3T) model for non-equilibrium thermochemistry of homogeneous chemical systems in their electronic ground state, derived using the multi-group maximum-entropy principle. The proposed framework captures both vibrational and rotational non-equilibrium effects, thereby enabling an accurate representation of thermochemical behavior in high-temperature gases. To account for departures from equilibrium dissociation, a non-equilibrium correction factor derived from ro-vibrational State-to-State (StS) analysis is incorporated into the population distribution. Because this correction is constructed from the quasi-steady-state (QSS) distribution, the resulting model is valid only for systems and conditions under which the QSS assumption holds. The proposed 1-bin 3T model is validated across a broad range of zero-dimensional test cases through comparisons with high-fidelity StS simulations and Direct Molecular Simulations (DMS). The results demonstrate excellent agreement between the reduced-order model and high-fidelity approaches for both dissociating and recombining environments. Overall, the 1-bin 3T model shows strong potential as an efficient and accurate framework for modeling non-equilibrium thermochemical behavior, while keeping computational costs orders of magnitude lower than the high fidelity ro-vibrational StS simulations.

\end{abstract}

\maketitle

\section{Introduction}
Modeling of hypersonic aerothermodynamics surrounding a space re-entry vehicle and hypersonic aircraft has been a topic of interest for decades. Hypersonic flows are characterized by strong shocks that produce substantial excitation of internal energy modes, molecular dissociation and ionization. The high altitude and high speed encountered in a typical hypersonic flight result in a combination of low collision frequency and low residence time. Therefore, in the absence of sufficient collisions among particles, non-equilibrium phenomenon becomes an important aspect of the hypersonic flow. Non-equilibrium effects play a critical role in determining the evolution of species composition and energy transfer processes which can significantly alter the shock stand-off distance, thereby affecting the surface heat flux and pressure distributions. Accurately capturing such non-equilibrium behavior is therefore indispensable for the development of thermochemical models that enable accurate prediction of hypersonic flowfields and, ultimately, the reliable design of hypersonic vehicles.

The non-equilibrium behavior can be simulated using thermochemical models which range from highly accurate high-fidelity State-to-State (StS) models \cite{Panesi2013_N3,Macdonald2020_N3,Sung_Min_N2O,Kim_Hydrogen,Boyd_emp_sts,Boyd_Sts_N2,Boyd_O2,Meuwly_JCP2014,Meuwly_JCP2015,Meuwly_PhysicalChem2017,Notey-Recomb-JCP2025,Panesi_Vibcoll_2009} to highly simplified but computationally inexpensive empirical models\cite{Adamovich,SSH_Approx,Zener,Park2T,MMT,Marrone1,Marrone2} like the widely used Park-2T model\cite{Park2T}. The StS models are the state of the art thermochemical models to capture the non-equilibrium behavior of high-speed reacting flows. These collisional models are based on ab-initio quantum chemistry data and use scattering methods such as the quasi-classical trajectory (QCT) calculations \cite{Schwenke1988_VVTC} to determine the kinetic processes among the internal energy levels of the species. Depending on the model resolution, the StS models track individual states (ro-vibrational or vibrational states) by treating them as a pseudo-species and hence are highly accurate but computationally demanding. Part of this cost stems from the strong non-linear coupling and stiffness of the governing source terms, which penalize the numerical integration of the thermochemical system. The dominant limitation, however, is the size of the system itself: every internal energy level becomes a pseudo-species, and the number of kinetic processes among these levels grows combinatorially with their number, so that the number of species and reactions explodes. As a consequence, StS models cannot be incorporated into three-dimensional flow solvers.

The computational limitations associated with the StS models bring forth the need to develop reduced-order models that are both accurate and computationally tractable for application to multi-dimensional CFD simulations. In particular, the development of non-equilibrium models requires a careful treatment of both non-equilibrium internal energy exchange and dissociation/recombination processes to ensure that the reduced system accurately reproduces the transient species evolution and thermal relaxation behavior. Significant work has been done in recent years focusing on the development of reduced-order models for non-equilibrium hypersonic flows to surpass the computational limitations of StS models while maintaining accuracy. Efforts have been made to improve empirical models such as the Park-2T \cite{Parkmodel} and Marrone-Treanor \cite{Marrone1, Marrone2} models by updating the model parameters obtained from experiments with those obtained from high-fidelity analysis of the chemical system under consideration \cite{MMT,Grover_comparative_2025,MMT-JTHT}. The non-Boltzmann distributions of internal energy levels have also been modeled in the past\cite{Singh_NonBoltz}. For instance, Singh et al. \cite{Singh_17_surprisal} employed an information-theoretical approach based on surprisal analysis of Direct Molecular Simulation (DMS) data to characterize the quasi-steady-state (QSS) rotational and vibrational energy distributions. More recently, data-driven and projection-based reduced-order models have received increasing attention for homogeneous reacting systems\cite{Zanardi_Ar, Zanardi_PG}. These approaches typically construct compact representations of the thermochemical state using reduced basis functions derived from simulation data. Data-driven methods can provide substantial computational acceleration; however, ensuring robustness, stability, and thermodynamic consistency remains a significant challenge, particularly for highly stiff nonequilibrium kinetics. A physics-derived model reduction strategy based on the maximum entropy principle and coarse-graining \cite{Liu2015,Macdonald_2019_CGQCT,Macdonald2018_CGQCT,Macdonald2018_CGQCT2,Sahai2017_Adaptive,Sharma2020_Quadratic,Venturi2020_O3Diss} has emerged as a promising pathway for modeling non-equilibrium flows. This technique relies on the chemistry data obtained from first principles to determine the macroscopic group coefficients. The main idea behind this model reduction is the grouping of internal energy levels into bins based on the criteria that all the energy states in a single bin are likely to equilibrate with each other. Hence, the group macroscopic properties are derived by assuming a Maxwell-Boltzmann equilibrium distribution of states within each group.

When the system is in QSS, an ideal coarse-grained closure would reproduce the QSS distribution even with a single group. This is not possible with the closure adopted by existing coarse-grained models (CGM): the Maxwell-Boltzmann distribution imposed within each group cannot reproduce the QSS population, which departs strongly from a Boltzmann distribution in the high-lying levels that drive dissociation. Previous studies have therefore shown that existing CGM require multiple bins to predict the macroscopic properties well \cite{Sahai2017_Adaptive,Venturi2020_O3Diss,2015Panesi}, at a higher computational cost  than that of their multi-temperature counterparts. For systems and conditions in which the gas is in QSS, the strategy adopted in this work is instead to retain a single group and to introduce a QSS correction to the intra-group distribution, so that the reduced model recovers the correct dissociation behavior at a cost comparable to that of the empirical multi-temperature models. To this end, this study presents an efficient physics-based one-bin, three-temperature (3T) framework grounded in the maximum entropy principle and derived from ab initio quantum chemistry databases, enabling its practical application in multi-dimensional Navier–Stokes CFD simulations. A non-equilibrium factor, derived from StS analysis, is used to model the quasi-steady state (QSS) dissociation phenomenon. The proposed model is validated using 0-D reactor cases through comparison with high-fidelity StS and DMS simulations. The 0-D simulations using the ro-vibrational collisional StS model and the 1-bin 3T model were performed using PLATO \cite{Munafo_JTHT_plato_2025}, the in-house thermochemical library. The 0-D reactor configurations provide an important framework for investigating the reduced-order formulation in a controlled setting before extension to multi-dimensional flow simulations. This analysis enables a focused assessment of the thermochemical model reduction by isolating the thermochemical behavior from additional transport-related complexities.

The paper structure is as follows: Section II describes the modeling methodology used in the derivation of the 1-bin 3T model. To begin with, Section II (A) introduces the multi-group maximum entropy (MGME) model, which is later used to derive the multi-temperature modeling for energy transfer and dissociation processes as shown in Sections II (B) and II (C), respectively. Section II (D) presents the expressions for macroscopic reaction rate and energy transfer coefficients for diatom-atom and diatom-diatom interactions. The potential energy surfaces used in the determination of kinetics and thermodynamics data for the reduced-order model are described in Section II (E). Section III presents the macroscopic quantities and comparative results of the proposed 1-bin 3T model with the high-fidelity StS and DMS simulations for a range of zero-dimensional test cases. Lastly, Section IV concludes the paper with a summary of the key takeaway points. 

\section{Physical Modeling}
This section outlines the reduction methodology used to derive macroscopic reaction rates and energy transfer coefficients by employing a non-equilibrium treatment of the internal energy mode by splitting it into its rotational and vibrational counterparts. A non-equilibrium correction factor accounting for QSS dissociation is also obtained using the StS/DMS data. Finally, the section presents the closure of the group conservation equations in the form of macroscopic coefficients obtained from the proposed model reduction.

\subsection{Multi-Group Maximum Entropy model}
The derivation of the proposed 1-bin 3T model leverages the Multi-Group Maximum Entropy (MGME) \cite{Liu2015} approach
to define the population distribution of rotational and vibrational energy levels. To introduce this concept, this sub-section presents the conventional coarse-grained model reduction approach derived from the maximum-entropy principle, which employs a linear form of reconstruction for the logarithm of population distribution, resulting in a Maxwell-Boltzmann distribution within individual bins, as described in detail by Liu et al \cite{Liu2015}. Therefore, the internal energy levels in a bin are distributed as:

\begin{equation}
    \frac{n_{vJ}^{M}}{{n}_{p}^{M}}=\frac{g_{vJ}^{M}exp\left(\frac{-E_{vJ}^{M}}{k_{B}T_{p}^{I,M}}\right)}{\sum_{i(v,J) \in I_{p,M}} g_{vJ}^{M}exp\left(\frac{-E_{vJ}^{M}}{k_{B}T_{p}^{I,M}}\right)},
\end{equation}

\noindent
where $n_{vJ}^{M}$, $g_{vJ}^{M}$, and $E_{vJ}^{M}$ are the number density, degeneracy and internal energy of level $i(v,J)$, of molecule M with vibrational quantum number $v$ and rotational quantum number $J$, respectively. Here, $n_{p}^{M}$ and $T_{p}^{I,M}$ represent the number density and the internal temperature of the bin $p$ for a molecule M, respectively, with $I_{p,M}$ being the set of all internal levels in the bin $p$ of molecule M and $k_{B}$ representing the Boltzmann constant. The macroscopic governing equations and hence the macroscopic reaction rate and energy transfer coefficients are obtained by taking moments of the master equations. Consequently, the rate of change of the number density ($n_{p}^{M}$) and internal energy ($e_{p}^{I,M}$) of the $p$-th bin are given by:

\begin{align}
\begin{split}
    \frac{dn_{p}^{M}}{dt}=\sum_{i(v,J)\in I_{p,M}}\frac{dn_{vJ}^{M}}{dt}, \quad 
    \frac{de_{p}^{I,M}}{dt}=\sum_{i(v,J)\in I_{p,M}}E_{vJ}^{M}\frac{dn_{vJ}^{M}}{dt}.
\end{split}
\end{align}

\subsection{Modeling of rotational and vibrational non-equilibrium}
In the MGME framework, the grouping of the internal energy levels is not imposed a priori but is itself an outcome of the reduction: the adaptive coarse-graining strategy of Sahai et al. \cite{Sahai2017_Adaptive} lets the group structure follow from the energy transfer and dissociation kinetics of the system. The conventional formulation models the evolution of the total internal energy of each group without distinguishing rotational and vibrational contributions. The present work targets conditions of mild non-equilibrium, for which the conventional separation of the internal energy into rotational and vibrational modes holds true, and therefore applies the MGME reduction to each mode separately, as previously described by Munaf\`o et al. \cite{Panesi_2015}. The modeling of rotational and vibrational modes separately is crucial to accurately capture the flow physics and internal energy evolution of the concerned chemical systems, especially at temperatures lower than 15000 K for oxygen and 25000 K for nitrogen, for which previous high-fidelity studies \cite{Panesi2013_N3, Grover_2019_oxygen} have shown separate relaxation of the two modes, with the rotational mode being much faster. Neglecting the separation of the two internal energy modes leads to an overestimation of the energy transfer rate, which can significantly affect the predicted flow features in multi-dimensional hypersonic flow simulations.

To that end, an expression for the ro-vibrational population distribution in terms of rotational and vibrational energies is derived by applying the MGME approach to the internal energy levels in such a manner as to account for both the rotational and vibrational non-equilibrium effects. The resulting expression aims to capture the non-Boltzmann distribution of ro-vibrational levels, which is crucial to determine accurate macroscopic group coefficients that are obtained by taking moments of master equations, as described in detail later in Section II.D.

Consider a molecule in its electronic ground state whose vibrational states are grouped into different bins. To account for the rotational non-equilibrium, let the rotational levels of each vibrational bin of molecule M be represented by a Maxwell-Boltzmann distribution at a single rotational temperature $T^{R,M}$. The number density of ro-vibrational levels is thus defined as \cite{Panesi_2015}:

\begin{equation} 
\label{Pop_rot}
\frac{n_{vJ}^{M}}{\tilde{n}_{v(M)}}=\frac{g_{vJ}^{M}exp\left(\frac{-\Delta E_{vJ}^{M}}{k_{B}T^{R,M}}\right)}{\sum_{J \in J_{v}(M)} g_{vJ}^{M}exp\left(\frac{-\Delta E_{vJ}^{M}}{k_{B}T^{R,M}}\right)},
\end{equation}

\noindent
where $J_{v}(M)$ is the set of rotational levels in a vibrational bin with vibrational quantum number $v$, $\Delta E_{vJ}^{M}$ is the rotational energy of an internal energy level $i(v,J)$ of molecule M and $\tilde{n}_{v(M)}$ is the number density of the bin pertaining to vibrational quantum number $v$ of molecule M. Next, consider a Maxwell-Boltzmann distribution at a vibrational temperature $T^{V,M}$ to describe the population of each vibrational bin of molecule M. The resulting population distribution is expressed as \cite{Panesi_2015}:

\begin{equation}
\label{pop_Vib}
    \frac{\tilde{n}_{v(M)}}{{n}_{M}}=\frac{\tilde{Q}_{v(M)}^{r}exp\left(\frac{-E_{v(M)}}{k_{B}T^{V,M}}\right)}{Q_{M}},
\end{equation}

\noindent where $\tilde{Q}_{v(M)}^{r}=\sum_{J \in J_{v}(M)} g_{vJ}^{M} exp\left(\frac{-\Delta E_{vJ}^{M}}{k_{B}T^{R,M}}\right)$ and $Q_{M}=\sum_{v \in I_{v(M)}}\tilde{Q}^{r}_{v(M)} exp\left(\frac{-E_{v(M)}}{k_{B}T^{V,M}}\right)$ with $I_{v(M)}$ being the set of all vibrational bins of molecule M and $E_{v(M)}$ represents the vibrational energy of the ro-vibrational level of molecule M with vibrational quantum number $v$. For the special case of single rotational and vibrational temperature, combining Eqs. \ref{Pop_rot} and \ref{pop_Vib} gives the population distribution function for a ro-vibrational energy level as:

\begin{equation}
    \label{Pop_distribution_function}
    \frac{n_{vJ}^{M}}{n_{M}}=\frac{g_{vJ}^{M}exp\left(\frac{-\Delta E_{vJ}^{M}}{k_{B}T^{R,M}}\right)exp\left(\frac{-E_{v(M)}}{k_{B}T^{V,M}}\right)}{\sum_{\substack{v \in I_{v}(M) \\ J \in J_{v}(M)}}g_{vJ}^{M}exp\left(\frac{-\Delta E_{vJ}^{M}}{k_{B}T^{R,M}}\right)exp\left(\frac{-E_{v(M)}}{k_{B}T^{V,M}}\right)}
\end{equation}

\subsection{Modeling of QSS dissociation}

In the coarse-grained modeling approach, the macroscopic dissociation rate coefficients are computed as weighted sums of the state-specific dissociation rate coefficients using the ro-vibrational population distribution, typically assumed to follow a Maxwell–Boltzmann equilibrium distribution. However, previous high-fidelity studies have shown that the non-equilibrium dissociation/recombination phenomenon is characterized by a quasi-steady state (QSS) distribution of the internal levels \cite{Panesi2013_N3, Venturi2020_O3Diss,Sung_Min_N2O,SungMin_23_JChemPhys,Grover_2019_oxygen}. When a gas is subjected to non-equilibrium conditions, dissociation/recombination and energy transfer processes balance each other to attain a QSS state in which the population distribution is time-invariant and deviates significantly from the Maxwell-Boltzmann distribution. For a single bin case, failure to model the dissociation process using this QSS distribution would result in a significant disparity with the high-fidelity results. This is primarily because the majority of the bulk dissociation/recombination occurs during the QSS period. To account for this behavior, the present work introduces a state-dependent non-equilibrium factor, which scales the population distribution expression described by Eq. \ref{Pop_distribution_function} to capture the deviation from equilibrium during the QSS period. Since the proposed functional form is identical for all the chemical systems considered, the methodology is presented in detail only for the $\mathrm{O_{2}+O}$ system for the sake of brevity.

\subsubsection{Non-equilibrium factors for diatom-atom chemical systems}
\begin{figure}
    \centering
    \includegraphics[width=1\linewidth]{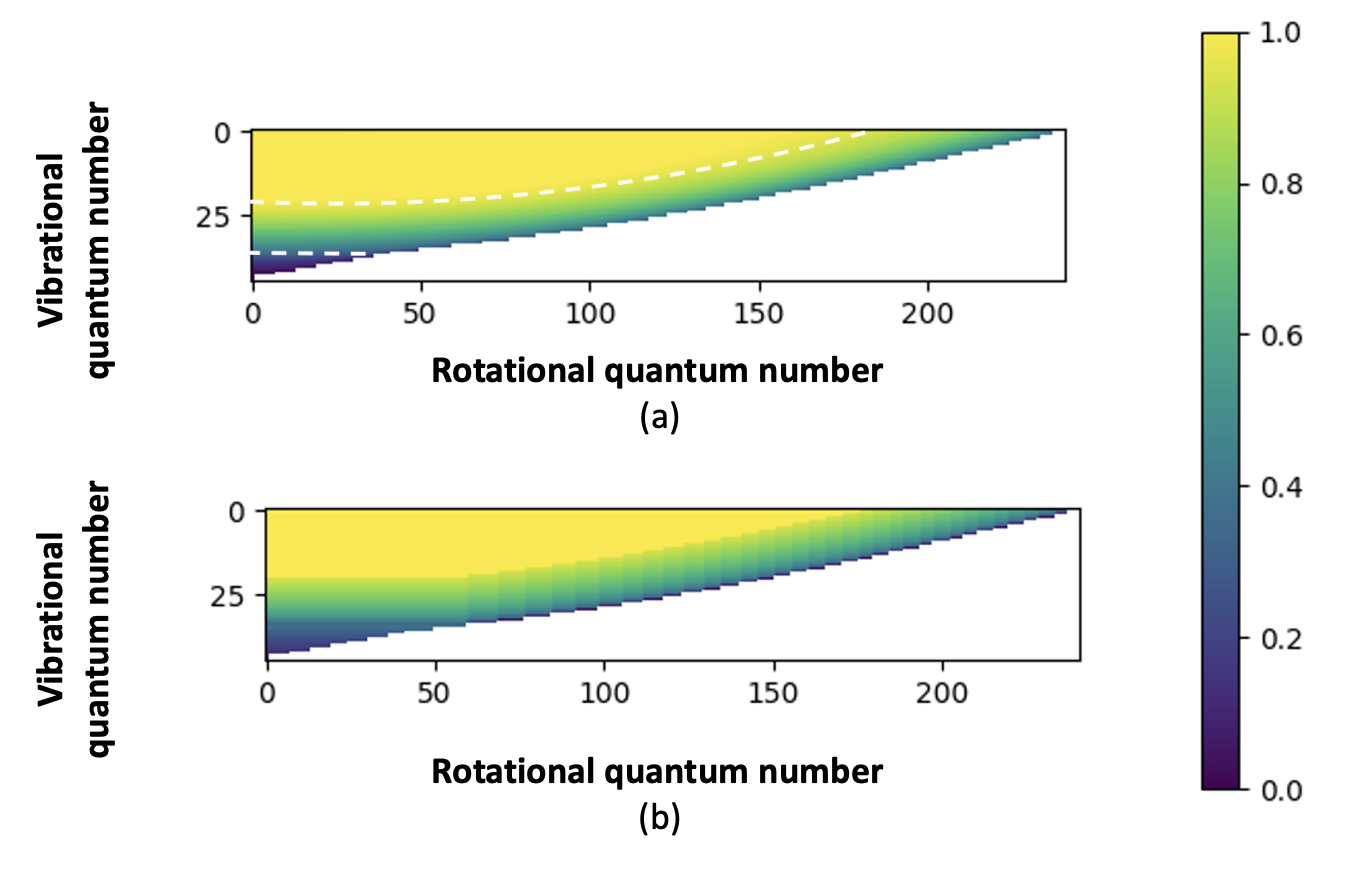}
    \caption{\textbf{Non-equilibrium population ratios obtained from (a) StS data and (b) Derived non-equilibrium factor during QSS at a heat bath temperature of 8000 K for $\mathrm{O_{2}+O}$ chemical system}}
    \label{fig: Non-equilibrium diss}
\end{figure}


The approach for determining the non-equilibrium factor is based on analyzing the StS QSS population distribution data to characterize the deviation from the equilibrium distribution \cite{Park}. In this work, a non-equilibrium population ratio is defined as the ratio between the actual (StS) population distribution and the equilibrium distribution (Eq. \ref{Pop_distribution_function}) at the corresponding QSS rotational and vibrational temperatures. The first step is to determine the non-equilibrium population ratios from StS data to qualitatively visualize the extent of non-equilibrium observed in each ro-vibrational level during QSS, as shown in Figure \ref{fig: Non-equilibrium diss} (a). A non-equilibrium population ratio of 1 indicates the levels being at equilibrium distribution at the QSS temperatures. A qualitative analysis of the figure shows that, essentially, the non-equilibrium behavior of internal energy levels can be roughly divided into three regions, as represented by white dashed lines. The plot shows that the levels with low rotational and vibrational quantum numbers have the least deviation from equilibrium, while the levels with low rotational and high vibrational quantum numbers ($\mathrm{v>35}$) show significant deviation due to their population ratio being significantly low. This observation is in-line with the assumption of preferential dissociation from high-lying vibrational states, thereby highlighting a strong correlation of the non-equilibrium behavior with the vibrational quantum number for energy levels with high vibrational energy. Additionally, a high correlation is observed between the non-equilibrium population ratio and the vibrational deficit for the intermediate vibrational quantum numbers. This is analogous to the observation made in the work by Venturi et al.\cite{Venturi2020_O3Diss}, which showed that the magnitude of the dissociation rate coefficients is correlated to the distance from the centrifugal barrier. In the present context, vibrational deficit of the ro-vibrational level $\mathrm{i(v,J)}$ of $O_{2}$, $\mathrm{v_{deficit,vJ}^{O_{2}}}$, is defined as:

\begin{equation}
    v_{deficit,vJ}^{O_{2}}=min(v_{max,J}^{O_{2}},35)-v(O_{2}),
\end{equation}

\noindent
where $v_{max, J}^{O_{2}}$ is the maximum vibrational quantum number pertaining to the rotational quantum number $J$ of the ro-vibrational level $i(v,J)$ of $O_{2}$.
    
    

\begin{figure}
    \centering
    \includegraphics[width=1\linewidth]{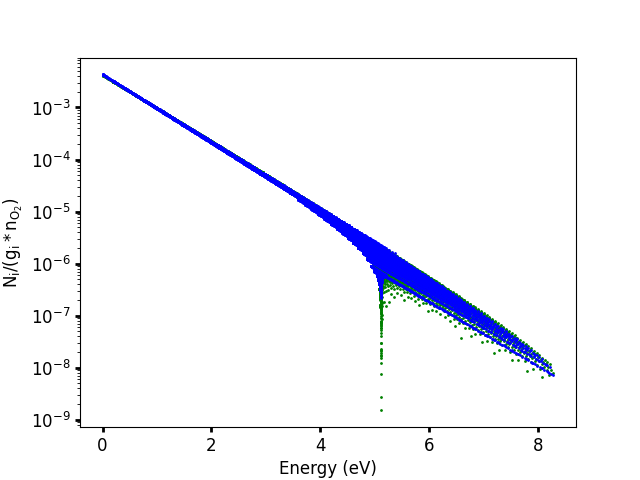}
    \caption{\textbf{Population distribution of internal energy levels of $O_{2}$ in QSS distribution at heat bath temperature of 8000 K where green indicates StS and blue indicates reconstructed distribution obtained from Eq.(\ref{QSS_pop}) }}
    \label{fig: QSS_pop}
\end{figure}
The non-equilibrium population ratio $R^{O_{2}}_{vJ}$ obtained from the 1-bin 3T model reduction is expressed as:
\begin{align}
\label{Pop ratio}
    R_{vJ}^{O_{2}}=\frac{n_{vJ,QSS}^{O_{2}}}{n_{O_{2}}} \times \frac{n_{O_{2}}}{n_{vJ}^{O_{2}}},
\end{align}
\noindent
The QSS distribution for the 1-bin 3T model is obtained by multiplying the population distribution expression in Eq. \ref{Pop_distribution_function} with a state-specific non-equilibrium factor, $F_{vJ}^{O_{2}-O}$, as follows:
\begin{align}
\label{QSS_pop}
\begin{split}
\frac{n_{vJ,QSS}^{O_{2}}}{{n}_{O_{2}}}=\frac{F_{vJ}^{O_{2}-O}g_{vJ}^{O_{2}}exp\left(\frac{-\Delta E_{vJ}^{O_{2}}}{k_{B}T^{R,O_{2}}}\right)exp\left(\frac{-E_{v(O_{2})}}{k_{B}T^{V,O_{2}}}\right)}{\sum_{\substack{v \in I_{v}(O_{2}) \\ J \in J_{v}(O_{2})}} F_{vJ}^{O_{2}-O}g_{vJ}^{O_{2}}exp\left(\frac{-\Delta E_{vJ}^{O_{2}}}{k_{B}T^{R,O_{2}}}\right)exp\left(\frac{-E_{v(O_{2})}}{k_{B}T^{V,O_{2}}}\right)}
\end{split}
\end{align}

\noindent and the ratio $n_{O_{2}}/n_{vJ}^{O_{2}}$ is obtained from Eq. \ref{Pop_distribution_function}.
Based on the insights obtained from Figure \ref{fig: Non-equilibrium diss} (a), the non-equilibrium factor, $F_{vJ}^{O_{2}-O}$, is obtained for each ro-vibrational level $i(v,J)$ by tuning the parameters to qualitatively reproduce the StS-derived non-equilibrium population ratios and QSS population distribution using Eqs.\ref{Pop ratio} and \ref{QSS_pop}, respectively. The resulting non-equilibrium factor is expressed as:

\begin{align}
F_{vJ}^{O_{2}-O}=\begin{cases}
    0.12\left(\frac{v_{max}^{O_{2}}}{v(O_{2})}\right)^{5}, & \text{if  } v(O_{2})\geq 35,\\
    1.4\left(\frac{v_{deficit,vJ}^{O_{2}}}{v_{max}^{O_{2}}}\right)^{0.37}, & \text{if } v(O_{2}) \text{<} 35
    \text{ and }v_{deficit,vJ}^{O_{2}}\text{<} 15,\\
    1, &\text{if }v(O_{2})\text{<} 35 \text{ and }v_{deficit,vJ}^{O_{2}}\geq 15. 
\end{cases}
\end{align}

\noindent
Here, $v_{max}^{O_{2}}$ is a constant normalizing factor equal to the value of maximum vibrational quantum number for the oxygen molecule, taken as 44. Figure \ref{fig: Non-equilibrium diss} (b) shows that the use of the derived non-equilibrium factor in Eq. \ref{Pop ratio} is able to closely reproduce the StS non-equilibrium population ratios. A qualitative comparison of the reconstructed population distribution obtained by using Eq. (\ref{QSS_pop}) is also in excellent agreement with the StS QSS distribution, as shown in Figure \ref{fig: QSS_pop}. The non-equilibrium factor is determined from the QSS distribution at a single reference heat-bath temperature (8000 K for $\mathrm{O_{2}+O}$) and is assumed to be independent of temperature in the present model; the same assumption is adopted for the $\mathrm{N_{2}+N}$, $\mathrm{O_{2}+O_{2}}$ and $\mathrm{N_{2}+N_{2}}$ factors presented below. The adequacy of this assumption is assessed in Section III through the comparison of the resulting QSS dissociation rate coefficients with DMS data over the full range of temperatures considered.

A similar StS analysis was performed on the kinetics database of $\mathrm{N_{2}+N}$ chemical system at T=10000 K to determine the non-equilibrium factor. The non-equilibrium factor, $F_{vJ}^{N_{2}-N}$, thus obtained, is expressed as:

\begin{align}
F_{vJ}^{N_{2}-N}=\begin{cases}
    0.137\left(\frac{v_{max}^{N_{2}}}{v(N_{2})}\right)^{8}, & \text{if  } v(N_{2})\geq 48\\
    1.28\left(\frac{v_{deficit,vJ}^{N_{2}}}{v_{max}^{N_{2}}}\right)^{0.27}, & \text{if } v(N_{2}) \text{<} 48
    \text{ and }v_{deficit,vJ}^{N_{2}}\text{<} 15\\
    1, &\text{if }v(N_{2})\text{<} 48 \text{ and }v_{deficit,vJ}^{N_{2}}\geq 15 
\end{cases}
\end{align}

\noindent where $v_{deficit,vJ}^{N_{2}}=min(v_{max,J}^{N_{2}},48)-v(N_{2})$, and $v_{max,J}^{N_{2}}$ is the maximum vibrational quantum number pertaining to the rotational quantum number $J$ of the ro-vibrational level $i(v,J)$ of $N_{2}$, and $v_{max}^{N_{2}}=54$.

\subsubsection{Non-equilibrium factors for diatom-diatom chemical systems}
\begin{figure}
    \centering
    \includegraphics[width=1\linewidth]{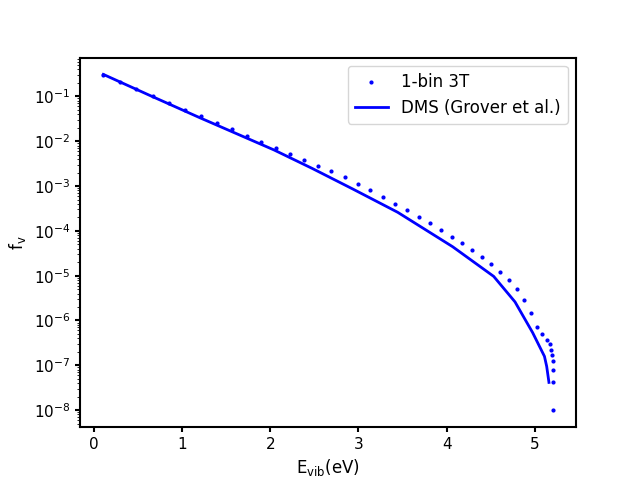}
    \caption{\textbf{Comparison of QSS vibrational energy distributions obtained using the 1-bin 3T model reduction with DMS \cite{Grover_2019_oxygen} for $\mathrm{O_{2}+O_{2}}$ chemical system at T=6000 K}}
    \label{fig: Vib_distr_O4}
\end{figure}

The functional form of the non-equilibrium factor for diatom-diatom interactions is considered the same as that for the corresponding diatom-atom interactions presented above. In the absence of availability of fully-resolved ro-vibrational StS kinetics data for diatom-diatom system, the non-equilibrium factor in this case is derived by fitting the QSS vibrational energy distributions obtained from DMS \cite{Grover_2019_oxygen} to that obtained from the 1-bin 3T model reduction, as shown in Figure \ref{fig: Vib_distr_O4} for a heat bath temperature of 6000 K for $\mathrm{O_{2}+O_{2}}$ chemical system.  The normalized vibrational energy distributions are obtained by summing the ro-vibrational population distribution over all the rotational levels pertaining to the given vibrational energy as follows:

\begin{equation}
    f_{v}=\frac{\sum_{J_{v}(O_{2})}n_{vJ,QSS}^{O_{2}}}{n_{O_{2}}}.
\end{equation}

The non-equilibrium factor proposed to account for the non-equilibrium behavior in the $\mathrm{O_{2}+O_{2}}$ system is as follows:
\begin{align}
F_{vJ}^{O_{2}-O_{2}}=\begin{cases}
    0.015\left(\frac{v_{max}^{O_{2}}}{v(O_{2})}\right)^{3}, & \text{if  } v(O_{2})\geq 35,\\
    2.68\left(\frac{v_{deficit,vJ}^{O_{2}}}{v_{max}^{O_{2}}}\right)^{1.25}, & \text{if } v(O_{2}) \text{<} 35
    \text{ and }v_{deficit,vJ}^{O_{2}}\text{<} 20,\\
    1, &\text{if }v(O_{2})\text{<} 35 \text{ and }v_{deficit,vJ}^{O_{2}}\geq 20.
\end{cases}
\end{align}

\noindent

The normalized vibrational energy distributions obtained from DMS \cite{Grover_N3N4_2022} for $\mathrm{N_{2}+N_{2}}$ system at T=20000 K were fitted by parametric tuning of the functional form similar to that proposed for $\mathrm{N_{2}+N}$ system. The resulting non-equilibrium factor, $\mathrm{F_{vJ}^{N_{2}-N_{2}}}$, is expressed as:

\begin{align}
F_{vJ}^{N_{2}-N_{2}}=\begin{cases}
    0.058\left(\frac{v_{max}^{N_{2}}}{v(N_{2})}\right)^{8}, & \text{if  } v(N_{2})\geq 48\\
    1.8\left(\frac{v_{deficit,vJ}^{N_{2}}}{v_{max}^{N_{2}}}\right)^{0.6}, & \text{if } v(N_{2}) \text{<} 48
    \text{ and }v_{deficit,vJ}^{N_{2}}\text{<} 17\\
    1, &\text{if }v(N_{2})\text{<} 48 \text{ and }v_{deficit,vJ}^{N_{2}}\geq 17 
\end{cases}
\end{align}

\subsection{Application of the three-temperature model reduction to homogeneous chemical systems}\label{sec:coeffs}
This section presents a general multi-group expression of macroscopic zeroth- and first-order reaction rate and energy transfer coefficients for diatom-atom and diatom-diatom interactions. The derived general coefficients can be readily reduced to the 1-bin three-temperature (3T) model formulation by assuming that all the ro-vibrational levels are grouped into a single bin, characterized by one rotational and one vibrational temperature.

\subsubsection{Diatom-atom interactions}
The 1-bin 3T model is applied to the StS kinetics database for diatom-atom interactions by taking moments of the microscopic master equations. The StS kinetic processes considered for the model reduction are:

\begin{enumerate}
    \item Dissociation : \ce{$M$($i$) + $A$ <=>[$\kappa_{i}^{d}$][$\kappa_{i}^{r}$] $A$ + $A$ + $A$}
    \item Excitation/de-excitation : \ce{$M$($i$) + $A$ <=>[$\kappa_{i,j}$][$\kappa_{j,i}$] $M$($j$) + $A$}
\end{enumerate}

\noindent Here, the symbols $M$ and $A$ represent molecular and atomic species, respectively, and $\kappa_{i}^{d}$, and $\kappa_{i}^{r}$ are the dissociation and recombination reaction rate coefficients, respectively, of ro-vibrational level $i$. $\kappa_{i,j}$ and $\kappa_{j,i}$ refer to the excitation and de-excitation reaction rate coefficients, respectively, if $i<j$.
The microscopic master equations governing the population of ro-vibrational internal energy levels in a 0-D isothermal reactor are:
\begin{equation}
\begin{split}
    \frac{dn_{i}^{M}}{dt} =\sum_{j\neq i}\left(-\kappa_{i,j}n_{i}^{M}n_{A}+\kappa_{j,i}n_{j}^{M}n_{A}\right)
    +\left(-\kappa_{i}^{d}n_{i}^{M}n_{A}+\kappa_{i}^{r}n_{A}^{3}\right).
\end{split}
\end{equation}

\noindent
The macroscopic conservation equations for group number density, rotational and vibrational energy of the bin $p$ are written as:

\begin{widetext}
\begin{equation}
\begin{split}
\label{Diatom-atom-iso1}
    \sum_{i \in I_{p}}\frac{dn_{i}^{M}}{dt}= \frac{dn_{p}^{M}}{dt} =\sum_{\substack{q \in I_{g} \\ q\neq p}}\left(-{}^{0}K_{p,q}n_{p}^{M}n_{A}+{}^{0}\Bar{K}_{p,q}n_{q}^{M}n_{A}\right) 
    +\left(-{}^{0}C_{p}n_{p}^{M}n_{A}+{}^{0}\Bar{C}_{p}n_{A}^{3}\right),
\end{split}
\end{equation}

\begin{equation}
\begin{split}
\label{Diatom-atom-iso2}
    \sum_{i \in I_{p}}E_{i}^{R,M}\frac{dn_{i}^{M}}{dt}=\frac{de_{p}^{R,M}}{dt} =\sum_{q \in I_{g}}\left(-{}^{1}K_{p,q}^{R}n_{p}^{M}n_{A}+{}^{1}\Bar{K}_{p,q}^{R}n_{q}^{M}n_{A}\right)
    +\left(-{}^{1}C_{p}^{R}n_{p}^{M}n_{A}+{}^{1}\Bar{C}_{p}^{R}n_{A}^{3}\right),
\end{split}
\end{equation}

\begin{equation}
\begin{split}
\label{Diatom-atom-iso3}
    \sum_{i \in I_{p}}E_{i}^{V,M}\frac{dn_{i}^{M}}{dt}=\frac{de_{p}^{V,M}}{dt} =\sum_{q \in I_{g}}\left(-{}^{1}K_{p,q}^{V}n_{p}^{M}n_{A}+{}^{1}\Bar{K}_{p,q}^{V}n_{q}^{M}n_{A}\right)
    +\left(-{}^{1}C_{p}^{V}n_{p}^{M}n_{A}+{}^{1}\Bar{C}_{p}^{V}n_{A}^{3}\right),
\end{split}
\end{equation}
\end{widetext}

\noindent
where ${}^{0}K_{p,q}$ and ${}^{0}\Bar{K}_{p,q}$ denote the zeroth-order reaction rate coefficients for the excitation/de-excitation processes, ${}^{0}C_{p}$ and ${}^{0}\Bar{C}_{p}$ denote the grouped reaction rate coefficients for dissociation and recombination, respectively. ${}^{1}K_{p,q}^{R,V}$ and ${}^{1}\Bar{K}_{p,q}^{R,V}$ denote the first-order rotational and vibrational energy transfer coefficients for the excitation/de-excitation processes and ${}^{1}C_{p}^{R,V}$ and ${}^{1}\Bar{C}_{p}^{R,V}$ denote the grouped rotational and vibrational energy transfer coefficients for dissociation and recombination, respectively. Here, $I_{p}$ is the set of all ro-vibrational levels contained in bin $p$ and $I_{g}$ is the set of all bins. 

The expression for group reaction rate and energy transfer coefficients obtained by taking moments of StS master equations are :\\

\begin{equation}
\begin{split}
\label{Diss_rate_coeff}
    {}^{0}C_{p}=
    \frac{\sum_{i \in I_{p}}F_{i}^{M-A} \kappa_{i}^{d}(T)g_{i}exp\left(\frac{-E_{i}^{R}}{k_{B}T_{p}^{R}}\right)exp\left(\frac{-E_{i}^{V}}{k_{B}T_{p}^{V}}\right)}{\sum_{i \in I_{p}} F_{i}^{M-A}g_{i}exp\left(\frac{-E_{i}^{R}}{k_{B}T_{p}^{R}}\right)exp\left(\frac{-E_{i}^{V}}{k_{B}T_{p}^{V}}\right)},
\end{split}
\end{equation}

\begin{equation}
\begin{split}
    {}^{1}C_{p}^{R,V}=
    \frac{\sum_{i \in I_{p}}F_{i}^{M-A}\kappa_{i}^{d}(T) g_{i}E_{i}^{R,V} exp\left(\frac{-E_{i}^{R}}{k_{B}T_{p}^{R}}\right)exp\left(\frac{-E_{i}^{V}}{k_{B}T_{p}^{V}}\right)}{\sum_{i \in I_{p}} F_{i}^{M-A} g_{i}exp\left(\frac{-E_{i}^{R}}{k_{B}T_{p}^{R}}\right)exp\left(\frac{-E_{i}^{V}}{k_{B}T_{p}^{V}}\right)},
\end{split}
\end{equation}

\begin{equation}
\begin{split}
    {}^{1}K_{p,q}^{R,V}=
    \frac{\sum_{i \in I_{p}}\sum_{j \in I_{q}}\kappa_{i,j}(T) g_{i}E_{i}^{R,V} exp\left(\frac{-E_{i}^{R}}{k_{B}T_{p}^{R}}\right)exp\left(\frac{-E_{i}^{V}}{k_{B}T_{p}^{V}}\right)}{\sum_{i \in I_{p}}g_{i}exp\left(\frac{-E_{i}^{R}}{k_{B}T_{p}^{R}}\right)exp\left(\frac{-E_{i}^{V}}{k_{B}T_{p}^{V}}\right)},
\end{split}
\end{equation}

\noindent
where $T$ is the translational temperature, $T_{p}^R$ and $T_{p}^V$ are the rotational and vibrational temperatures, respectively, of group $p$. The reverse rate coefficients are obtained by applying the principle of detailed balancing at the microscopic level:

\begin{widetext}
\begin{equation}
    \begin{split}
     {}^{0}\Bar{C}_{p}=\frac{Q_{M}^{p} (T)Q_{M}^{trans} }{\left[g_{A}Q_{A}^{trans}\right]^{2}exp\left(\frac{-2E_{A}}{k_{B}T}\right)}  \frac{\sum_{i \in I_{p}}F_{i}^{M-A}\kappa_{i}^{d}(T) g_{i}exp\left(\frac{-\left(E_{i}^{R}+E_{i}^{V}\right)}{k_{B}T}\right)}{\sum_{i \in I_{p}} F_{i}^{M-A}g_{i}exp\left(\frac{-\left(E_{i}^{R}+E_{i}^{V}\right)}{k_{B}T}\right)}, 
    \end{split}
\end{equation}

\begin{equation}
    \begin{split}
     {}^{1}\Bar{C}_{p}^{R,V}=\frac{Q_{M}^{p}(T) Q_{M}^{trans} }{\left[g_{A}Q_{A}^{trans}\right]^{2}exp\left(\frac{-2E_{A}}{k_{B}T}\right)}\ \frac{\sum_{i \in I_{p}}F_{i}^{M-A}\kappa_{i}^{d}(T) g_{i} E_{i}^{R,V}exp\left(\frac{-\left(E_{i}^{R}+E_{i}^{V}\right)}{k_{B}T}\right)}{\sum_{i \in I_{p}} F_{i}^{M-A}g_{i}exp\left(\frac{-\left(E_{i}^{R}+E_{i}^{V}\right)}{k_{B}T}\right)}, 
    \end{split}
\end{equation}

\begin{equation}
\color{black}
    \begin{split}
        {}^{1}\Bar{K}_{p,q}^{R,V}=\frac{Q_{M}^{p}(T)}{Q_{M}^{q}(T_{q}^{R},T_{q}^{V})}\sum_{i \in I_{p}}\sum_{j \in I_{q}}\kappa_{i,j} \frac{g_{i}E_{i}^{R,V}exp\left(\frac{-\left(E_{i}^{R}+E_{i}^{V}\right)}{k_{B}T}\right)}{\sum_{i' \in I_{p}} g_{i'}exp\left(\frac{-\left(E_{i'}^{R}+E_{i'}^{V}\right)}{k_{B}T}\right)}
        \times \frac{exp\left(-\frac{E_{j}^{R}}{k_{B}T_{q}^{R}}-\frac{E_{j}^{V}}{k_{B}T_{q}^{V}}\right)}{exp\left(\frac{-\left(E_{j}^{R}+E_{j}^{V}\right)}{k_{B}T}\right)}.
    \end{split}
\end{equation}

\end{widetext}
\noindent where $Q_{M}^{p}$ is the group partition function for bin $p$, and $Q_{M,A}^{trans}$ are the translational partition functions for molecule $M$ and atom $A$, respectively. $E_{A}$ is the formation energy of atom $A$, such that $2E_{A}$ equals the dissociation energy of the molecule measured from its ground ro-vibrational level, and 
$g_{A}$ is the total degeneracy of the atomic ground state, including the nuclear-spin factor whenever it is included in the molecular level degeneracies $g_{i}$ (\textit{e.g.}, $g_{N}=4\times 3=12$ for nitrogen, since $g_{i}$ of $\mathrm{N_{2}}$ contains the nuclear-spin factor). These partition functions are expressed as:

\begin{equation}
    Q_{M}^{p}=\sum_{i \in I_{p}}g_{i}exp\left(\frac{-E_{i}}{k_{B}T}\right)
\end{equation}
where $E_{i}=E_{i}^{R}+E_{i}^{V}$, and, for a group with rotational and vibrational temperatures $T_{p}^{R}$ and $T_{p}^{V}$, $Q_{M}^{p}(T_{p}^{R},T_{p}^{V})=\sum_{i \in I_{p}}g_{i}\exp\left(-\frac{E_{i}^{R}}{k_{B}T_{p}^{R}}-\frac{E_{i}^{V}}{k_{B}T_{p}^{V}}\right)$, so that $Q_{M}^{p}(T)\equiv Q_{M}^{p}(T,T)$.

\begin{equation}
    Q_{M,A}^{trans}=\left(\frac{2 \pi k_{B}m_{M,A}T}{h_{P}^{2}}\right)^{3/2}
\end{equation}
where $h_{P}$ stands for Planck's constant and $m_{M,A}$ is the mass of molecular species $M$ or atomic species $A$. It should be noted that the recombination coefficient ${}^{0}\Bar{C}_{p}$ is the dissociation coefficient evaluated at $T_{p}^{R}=T_{p}^{V}=T$ divided by the equilibrium constant, \emph{i.e.}, ${}^{0}\Bar{C}_{p}={}^{0}C_{p}(T,T,T)/K_{eq}(T)$, so that the non-equilibrium factor $F_{i}^{M-A}$ enters the forward and reverse coefficients consistently and the net reaction rate vanishes at the equilibrium composition. Here, $K_{eq}(T)=\left[g_{A}Q_{A}^{trans}\right]^{2}\exp\left(-2E_{A}/k_{B}T\right)/\left[Q_{M}^{p}(T)\,Q_{M}^{trans}\right]$ is the equilibrium constant of the dissociation reaction; for the single-bin model, $Q_{M}^{p}(T)$ is the full internal partition function of the molecule.

It must be noted that this work proposes a single bin model, therefore $p=q=1$. For the special case of a single bin model, the zero-order excitation rate coefficients, ${}^{0}K_{p,q}$ and ${}^{0}\Bar{K}_{p,q}$ 
are excluded from Eq.~(\ref{Diatom-atom-iso1}) because the energy transitions within a single bin do not affect the number density of the bin. The first-order intra-bin coefficients ${}^{1}K_{p,p}^{R,V}$ and ${}^{1}\Bar{K}_{p,p}^{R,V}$, on the other hand, do not vanish: transitions between levels of the same bin redistribute energy between the rotational, vibrational and translational modes even though they leave $n_{p}^{M}$ unchanged. For $p=q=1$ these intra-bin terms are the only rotational and vibrational relaxation terms in Eqs. \ref{Diatom-atom-iso2} and \ref{Diatom-atom-iso3}, which is why the sums in those equations run over all groups, including $q=p$, whereas the sum in Eq. \ref{Diatom-atom-iso1} excludes it.

\subsubsection{Diatom-diatom interactions}
A model reduction approach similar to diatom-atom interactions would not be feasible for the diatom-diatom collisions due to a large number of possible transitions, thereby making it computationally intractable to calculate the full resolution of ro-vibrational StS kinetics a priori. Therefore, a coupling between the 1-bin 3T model and QCT is used, which bypasses the need for StS kinetics data and applies the model sampling of ro-vibrational levels while running scattering calculations. The temperature conditions used to sample the rovibrational levels in the QCT simulations, performed over a collision-energy range of 0.00367–1.10248 Hartree, are detailed in Sec.~\ref{sec:qss_coeffs}. The QCT details and statistical uncertainties associated with the computed macroscopic coefficients, quantified in terms of their standard deviation, are presented in Appendix A.


The possible transitions in diatom-diatom interactions for groups $p$, $q$, $r$ and $s$ are:

\begin{itemize}
\item Double dissociation: Both the molecules dissociate

\begin{center}
\ce{$M$($p$) + $M$($q$) <=>[${}^{0}C\sb{pq}^{dd}$][${}^{0}\Bar{C}\sb{pq}^{dd}$] $A$ + $A$ + $A$ + $A$}
\end{center}

\item Single dissociation: One of the molecules dissociates.

\begin{center}
\ce{$M$($p$) + $M$($q$) <=>[${}^{0}C\sb{pq,r}^{d1}$][${}^{0}\Bar{C}\sb{pq,r}^{d1}$] $A$ + $A$ + $M$($r$)}
\end{center}

\begin{center}
\ce{$M$($p$) + $M$($q$) <=>[${}^{0}C\sb{pq,r}^{d2}$][${}^{0}\Bar{C}\sb{pq,r}^{d2}$] $M$($r$) + $A$ + $A$}
\end{center}

\item Energy transfer: This includes inelastic and exchange energy transfer processes

\begin{center}
\ce{$M$($p$) + $M$($q$) <=>[${}^{0}K\sb{pq,rs}$][${}^{0}\Bar{K}\sb{pq,rs}$] $M$($r$) + $M$($s$)}
\end{center}

\end{itemize}

The group mass, rotational and vibrational energy conservation equations corresponding to the above mentioned transitions are as follows:
\begin{widetext}
\begin{equation}
\color{black}
\begin{split}
    \frac{dn_{p}^{M}}{dt} &=\sum_{q \in I_{g}}\bigl(-{}^{0}C_{pq}^{dd}n_{p}^{M}n_{q}^{M}+{}^{0}\Bar{C}_{pq}^{dd}n_{A}^{4}\bigr) +\sum_{q \in I_{g}} \sum_{\substack{r \in I_{g} \\ r \neq p}}\bigl(-{}^{0}C_{pq,r}^{d1}n_{p}^{M}n_{q}^{M}+{}^{0}\Bar{C}_{pq,r}^{d1}n_{A}^{2}n_{r}^{M}\bigr)+\sum_{q \in I_{g}} \sum_{\substack{r \in I_{g} \\ r \neq p}}\bigl(-{}^{0}C_{pq,r}^{d2}n_{p}^{M}n_{q}^{M}+{}^{0}\Bar{C}_{pq,r}^{d2}n_{A}^{2}n_{r}^{M}\bigr)\\&  \hspace{5cm}   +{\sum_{q \in I_{g}}\sum_{r \in I_{g}}\sum_{s \in I_{g}}}\left(1+\delta_{pq}-\delta_{pr}-\delta_{ps}\right){\left(-{}^{0}K_{pq,rs}n_{p}^{M}n_{q}^{M}+
{}^{0}\Bar{K}_{pq,rs}n_{r}^{M}n_{s}^{M}\right)},
\end{split}
\end{equation}

\begin{equation}
\color{black}
\begin{split}
    \frac{de_{p}^{R,M}}{dt} &=\sum_{q \in I_{g}}\bigl(-{}^{1}C_{pq}^{dd,R}n_{p}^{M}n_{q}^{M}+{}^{1}\Bar{C}_{pq}^{dd,R}n_{A}^{4}\bigr)+\sum_{q \in I_{g}}\sum_{r \in I_{g}}\bigl(-{}^{1}C_{pq,r}^{d1,R}n_{p}^{M}n_{q}^{M}+{}^{1}\Bar{C}_{pq,r}^{d1,R}n_{A}^{2}n_{r}^{M}-{}^{1}C_{pq,r}^{d2,R}n_{p}^{M}n_{q}^{M}+{}^{1}\Bar{C}_{pq,r}^{d2,R}n_{A}^{2}n_{r}^{M}\bigr)\\& \hspace{7cm} +\sum_{q \in I_{g}}\sum_{r \in I_{g}}\sum_{s \in I_{g}}\left(-{}^{1}K_{pq,rs}^{R}n_{p}^{M}n_{q}^{M}+{}^{1}\Bar{K}_{pq,rs}^{R}n_{r}^{M}n_{s}^{M}\right),
\end{split}
\end{equation}
\begin{equation}
\color{black}
\begin{split}
    \frac{de_{p}^{V,M}}{dt} &=\sum_{q \in I_{g}}\bigl(-{}^{1}C_{pq}^{dd,V}n_{p}^{M}n_{q}^{M}+{}^{1}\Bar{C}_{pq}^{dd,V}n_{A}^{4}\bigr)+\sum_{q \in I_{g}}\sum_{r \in I_{g}}\bigl(-{}^{1}C_{pq\chk{,r}}^{d1,V}n_{p}^{M}n_{q}^{M}+{}^{1}\Bar{C}_{pq,r}^{d1,V}n_{A}^{2}n_{r}^{M}-{}^{1}C_{pq,r}^{d2,V}n_{p}^{M}n_{q}^{M}+{}^{1}\Bar{C}_{pq,r}^{d2,V}n_{A}^{2}n_{r}^{M}\bigr)\\& \hspace{7cm}+\sum_{q \in I_{g}}\sum_{r \in I_{g}}\sum_{s \in I_{g}}\left(-{}^{1}K_{pq,rs}^{V}n_{p}^{M}n_{q}^{M}+{}^{1}\Bar{K}_{pq,rs}^{V}n_{r}^{M}n_{s}^{M}\right).
\end{split}
\end{equation}
\end{widetext}
\noindent
where $\delta_{xy}$ is the Kronecker delta function.
Here, $r$ denotes the group of the molecule that survives a single-dissociation event, and the sums over $q$, $r$ and $s$ in the energy transfer terms run over all groups, including $p$ itself, since the above equations are the moments of the ro-vibrational master equations and every ordered channel contributes once; for the single-bin model considered here, $p=q=r=s=1$. The present work excludes double-dissociation reactions as their contribution was found to be atleast an order of magnitude smaller when compared to single dissociation reactions. The macroscopic dissociation reaction rate and energy transfer coefficients are weighted averages of the microscopic reaction rate constants as written below:

\begin{align}
\label{Diss_ratecoeff_diatom-diatom}
{}^{0}C_{pq,r}^{x}=\frac{\sum_{i\in p} \sum_{j\in q} \sum_{k\in r}\kappa_{ij,k}^{x}  n_{i}^M n_{j}^M}{n_{p}^{M} n_{q}^{M}},
\end{align}

\begin{align}
{}^{1}C_{pq,r}^{x,(R,V)}=\frac{\sum_{i\in p} \sum_{j\in q} \sum_{k \in r}\kappa_{ij,k}^{x} E_{i}^{R,V} n_{i}^M n_{j}^M}{n_{p}^{M} n_{q}^{M}}.
\end{align}

\noindent
where $x$ represents $d1$ or $d2$ reactions and $\kappa_{ij,k}$  is the microscopic dissociation rate constant pertaining to internal levels $i$ and $j$ of the two colliding molecules with the product molecule in energy level $k$. 

Similar expressions for reaction rates and energy transfer coefficients for excitation processes were derived. Consider a microscopic (\emph{i.e.}, ungrouped) energy transfer reaction:
\begin{center}
\ce{$M$($i$) + $M$($j$) <=>[$\kappa\sb{ij,kl}$][$\kappa\sb{kl,ij}$] $M$($k$) + $M$($l$)}
\end{center}

\noindent
Then, the macroscopic (\emph{i.e.}, grouped) forward reaction and energy transfer coefficients for excitation are:

\begin{align}
    {}^{0}K_{pq,rs}=\sum_{i \in p}\sum_{j \in q}\sum_{k \in r}\sum_{l \in s}\frac{\kappa_{ij,kl}n_{i}^M n_{j}^M} {n_{p}^{M} n_{q}^{M}},
\end{align}

\begin{align}
    {}^{1}K_{pq,rs}^{R,V}=\sum_{i \in p}\sum_{j \in q}\sum_{k \in r}\sum_{l \in s}\frac{\kappa_{ij,kl} E_{i}^{R,V} n_{i}^M n_{j}^M} {n_{p}^{M} n_{q}^{M}}.
\end{align}

\begin{widetext}
\begin{equation}
\label{Rev_energytransfer}
    \begin{split}
        {}^{1}\Bar{K}_{pq,rs}^{R,V}=&\sum_{i \in I_{p}}\sum_{j \in I_{q}}\sum_{k \in I_{r}}\sum_{l \in I_{s}}\frac{\kappa_{ij,kl} E_{i}^{R,V} g_{i}exp\left(\frac{-\left(E_{i}^{R}+E_{i}^{V}\right)}{k_{B}T}\right)g_{j}exp\left(\frac{-\left(E_{j}^{R}+E_{j}^{V}\right)}{k_{B}T}\right)}{\sum_{i' \in I_{p}} g_{i'}exp\left(\frac{-\left(E_{i'}^{R}+E_{i'}^{V}\right)}{k_{B}T}\right)\sum_{j' \in I_{q}} g_{j'}exp\left(\frac{-\left(E_{j'}^{R}+E_{j'}^{V}\right)}{k_{B}T}\right)}\\
        & \times \frac{Q_{M}^{p}(T)Q_{M}^{q}(T)}{Q_{M}^{r}(T_{r}^{R},T_{r}^{V})Q_{M}^{s}(T_{s}^{R},T_{s}^{V})} \frac{exp\left(-\frac{E_{k}^{R}}{k_{B}T_{r}^{R}}-\frac{E_{k}^{V}}{k_{B}T_{r}^{V}}\right)exp\left(-\frac{E_{l}^{R}}{k_{B}T_{s}^{R}}-\frac{E_{l}^{V}}{k_{B}T_{s}^{V}}\right)}{exp\left(\frac{-\left(E_{k}^{R}+E_{k}^{V}\right)}{k_{B}T}\right)exp\left(\frac{-\left(E_{l}^{R}+E_{l}^{V}\right)}{k_{B}T}\right)}
        \end{split}
\end{equation}
\end{widetext}

The reaction rate and energy transfer coefficients are further expanded by substituting the expression for $n_{i}^{M}/n_{g}^{M}$ using Equations \ref{Pop_distribution_function} and \ref{QSS_pop} for excitation and dissociation processes, respectively. However, as mentioned before for diatom-atom interactions, for a 1-bin case, the zeroth-order forward and reverse reaction rate coefficients are not considered as the transitions within the same group do not alter the number density of the group.
In this work, exothermic trajectories were used to calculate the forward energy transfer coefficients for excitation processes because of better statistical convergence than the endothermic trajectories. The reverse energy transfer coefficients were then derived by applying detailed balancing. The resulting expression for the reverse energy transfer rate coefficients is given by Eq. \ref{Rev_energytransfer} which shows that the ro-vibrational energy levels must be sampled at translational temperature for determination of reverse energy transfer coefficient. The scaling factors are then multiplied during the post-processing step. An expression similar to Eq. \ref{Rev_energytransfer} can be derived for the recombination reaction rate and energy transfer coefficients. It must also be noted that the model reduction on diatom-diatom system was tested against DMS calculations which do not simulate recombination. Therefore, the present work does not consider recombination for diatom-diatom while it is considered when comparing reduced-order model results with StS kinetics for diatom-atom system.

\subsection{Potential energy surfaces}
As described in Sec.~\ref{sec:coeffs}, the macroscopic reaction rate and energy transfer coefficients are obtained from QCT derived StS kinetics data for the diatom-atom collisions and by coupling the 1-bin 3T method with scattering calculations for the diatom-diatom systems. The QCT method solves Hamilton's equations of motion to determine the dynamics of molecules and atoms during collisions. The QCT calculations require a potential energy surface (PES) to describe the energy of a system in terms of position of the atoms in order to solve the equations of motion. A potential energy surface is obtained by solving the electronic Schrödinger equation for many fixed nuclear geometries and using the resulting electronic energies as a multi-dimensional function of the nuclear coordinates. The in-house Fortran code, CoarseAIR \cite{Venturi2020_O3Diss,Venturi2020_O3Bayesian}, which is based on the original vectorized variable step size trajectory code (VVTC) developed by Schwenke \cite{Schwenke1988_VVTC} at NASA Ames Center, was used to perform the QCT calculations. 

To determine the StS kinetics for $\mathrm{O_{2}+O}$, a set of nine potential energy surfaces based on the extended multi-state complete-active-state second order perturbation theory (XMS-CASPT2) describing the electronically adiabatic collisions of a ground state triplet oxygen molecule with a ground state triplet oxygen atom developed by Varga et al.\cite{Varga_O3_PES} were used. Each of the nine PESs was given a statistical weight of 1/27, 3/27 and 5/27 for the singlet, triplet and quintet states, respectively. 
$\mathrm{O_{2}+O_{2}}$ collisions were simulated using the potential energy surfaces for singlet, triplet and quintet states developed by Paukku et al.\cite{Paukku_O4_PES,Paukku_O4_triplet_2018} Neglecting spin-orbit coupling, the statistical weights for the singlet, triplet and quintet states are 1/9, 1/3 and 5/9, respectively. The potential energy surfaces are derived using multi-state complete active space second order perturbation theory (MS-CASPT2). The ground state oxygen molecule has a total of 6115 ro-vibrational energy levels which includes both bound and quasi-bound levels. The dissociation energy of the oxygen molecule is 5.113 eV with 45 vibrational states and a maximum rotational quantum number of 240. The rules of quantum mechanics dictate a nuclear spin degeneracy of 0 and 1 for even and odd rotational quantum numbers, respectively, for molecular oxygen. Therefore, ro-vibrational levels with even rotational quantum number are forbidden. However, high temperature limit approximation applied to the rotational partition function allows the consideration of all rotational states by incorporating a degeneracy of $1/2$ for each ro-vibrational energy level \cite{Atkins}. The ro-vibrational degeneracy of oxygen molecule is represented as $\mathrm{g_{vJ}=g_{e}(2J+1)/2}$, where $g_{e}=3$ is the electronic state degeneracy.\cite{Venturi2020_O3Diss}

The StS kinetics database for $\mathrm{N_{2}+N}$ was obtained by performing QCT calculations on the ab initio $\mathrm{N_{4}}$ PES of Paukku et al. \cite{Paukku_13,Bender_2015} developed at the University of Minnesota by treating one atom to be far away. The database includes 9165 ro-vibrational internal energy levels with maximum vibrational and rotational quantum numbers of 54 and 281, respectively. The degeneracy of the internal energy levels is given by $g_{vJ}=g_{vJ}^{NS}(2J(i)+1)$, where $g_{vJ}^{NS}$ is the nuclear spin degeneracy of the nitrogen molecule and $J(i)$ is the rotational quantum number of level $i=i(v,J)$. The nuclear spin degeneracy equals 6 for levels with even $J(i)$ and 3 for levels with odd $J(i)$. The same PES was used to determine macroscopic rate coefficients for $\mathrm{N_{2}+N_{2}}$ interactions by sampling the energy levels using the 1-bin 3T model reduction while performing scattering calculations.

\section{Results}

This section presents the comparison of macroscopic quantities obtained by the application of the 1-bin 3T model reduction with high-fidelity ro-vibrational StS and DMS simulations for a range of test conditions. The thermochemical behavior of only the diatom-atom interactions is validated by comparison against the high-fidelity StS simulations. DMS simulations are used as a benchmark to validate the thermochemistry in pure mixtures involving both diatom-atom and diatom-diatom interactions.

\begin{figure*}
\centering
    
    \subfigure[$\mathbf{N_{2}+N}$]{\label{fig:QSS_N3}\includegraphics[width=0.47\textwidth]{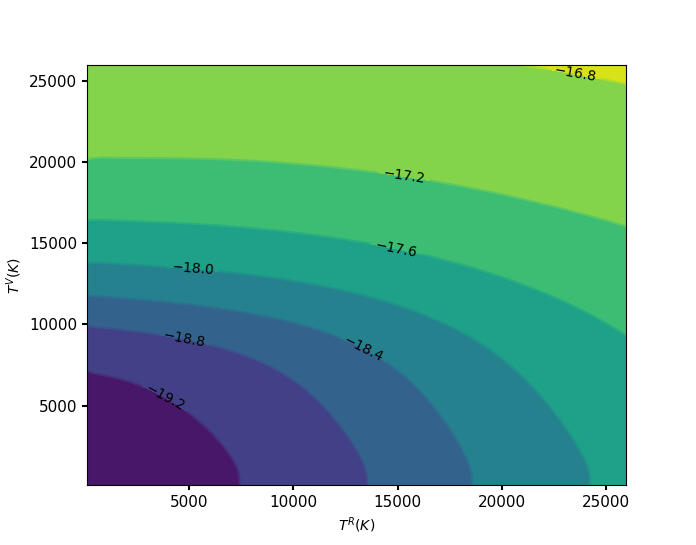}}   
    \subfigure[$\mathbf{N_{2}+N_{2}}$]{\label{fig:QSS_N4}\includegraphics[width=0.475\textwidth]{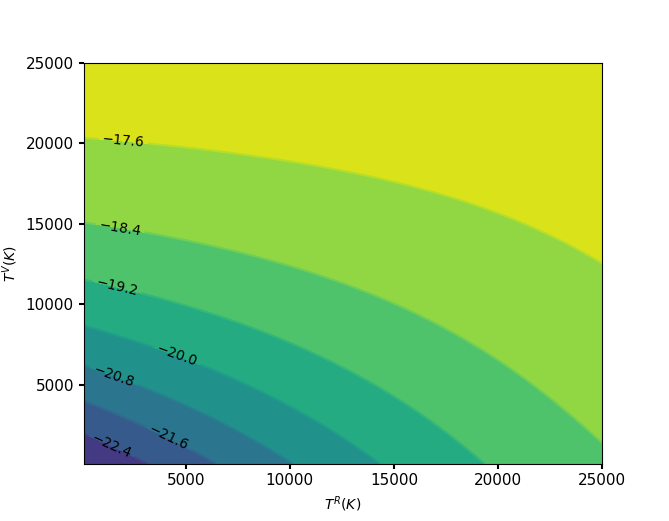}}
    \caption{\textbf{QSS dissociation reaction rate coefficients derived from the 1-bin 3T model at T=17500 K $\mathbf{(log_{10} {}^{0}C}$ $\mathbf{(m^{3}/s))}$}}
    \label{fig:QSS_contour}
\end{figure*}

\begin{figure*}
\centering
    
    \subfigure[$\mathbf{N_{2}+N}$]{\label{fig:Rot_coeff_N3}\includegraphics[width=0.49\textwidth]{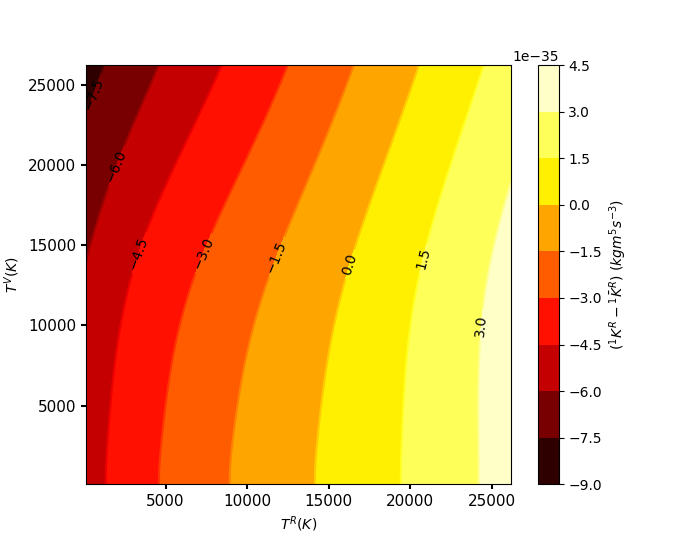}}   
    \subfigure[$\mathbf{N_{2}+N_{2}}$]{\label{fig:Rot_coeff_N4}\includegraphics[width=0.49\textwidth]{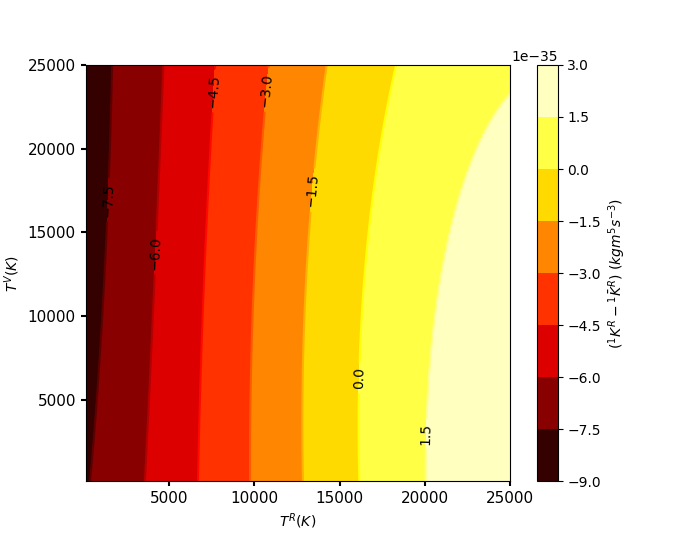}}
    \caption{\textbf{$(\mathbf{{}^{1}K^{R}-{}^{1}\Bar{K}^{R}})$ values at T=17500 K}}
    \label{fig:Figure2_Rot}
\end{figure*}

\subsection{Three-temperature QSS dissociation rate and energy transfer coefficients obtained from the 1-bin 3T model}\label{sec:qss_coeffs}
\begin{figure*}
\centering
    
    \subfigure[$\mathbf{N_{2}+N}$]{\label{fig:Vib_coeff_N3}\includegraphics[width=0.49\textwidth]{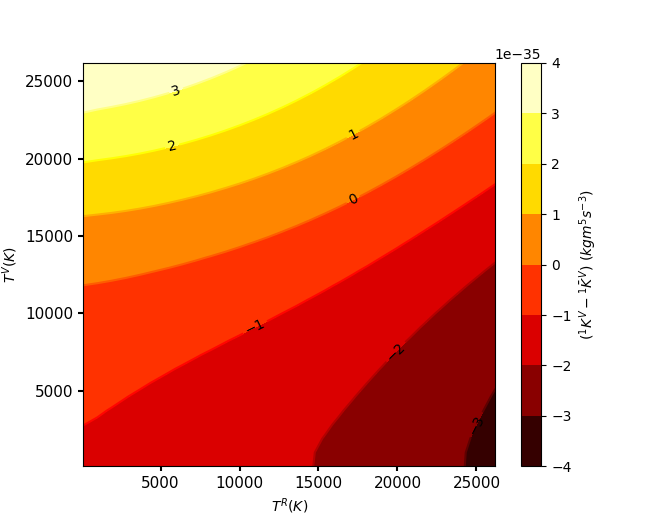}}   
    \subfigure[$\mathbf{N_{2}+N_{2}}$]{\label{fig:Vib_coeff_N4}\includegraphics[width=0.49\textwidth]{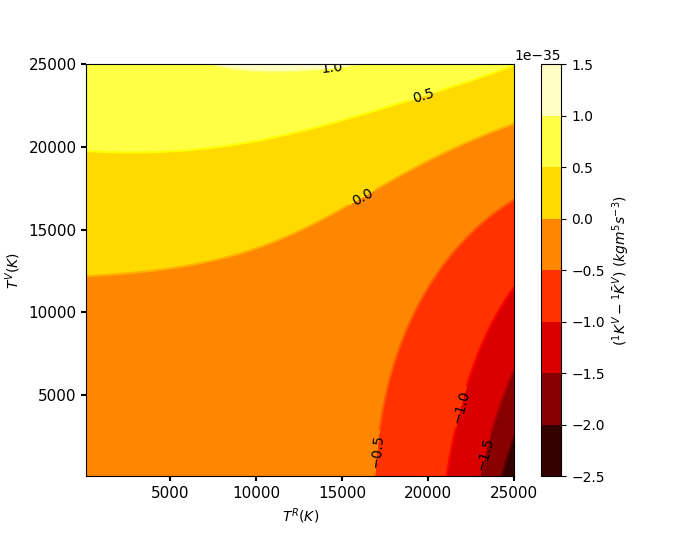}}
    \caption{\textbf{$(\mathbf{{}^{1}K^{V}-{}^{1}\Bar{K}^{V}})$ values at T=17500 K}}
    \label{fig:Figure3_Vib}
\end{figure*}

The macroscopic reaction rate and energy transfer coefficients obtained from the three-temperature model reduction are functions of translational, rotational, and vibrational temperatures, resulting in a three-dimensional table of numerical values. Due to the availability of ro-vibrational StS kinetics data for diatom-atom interactions, the group tables can be obtained for as many temperature points as required. Linear interpolations in the log-scale are used to determine the intermediate coefficient values. However, limited data points are available for the diatom-diatom interactions due to high computational cost associated with the QCT runs. For instance, generating a table for three rotational and three vibrational temperatures requires nine QCT runs. The diatom-diatom system results presented in this work are obtained by running model reduction-coupled QCT runs for rotational and vibrational temperatures of 5000 K, 7500 K and 10000 K for $\mathrm{O_{2}+O_{2}}$ and temperatures of 6000 K, 10000 K, 17500 K, and 25000 K for $\mathrm{N_{2}+N_{2}}$ chemical system. For each fixed translational temperature, the calculated data points at a given rotational temperature are fitted with a quadratic polynomial in logarithmic space as a function of vibrational temperature. This polynomial is then used to generate a fine grid of hundreds of temperature points over the desired temperature range, including both interpolated values within the range of the calculated data and extrapolated values beyond it. The intermediate rate coefficients are obtained by 3D linear interpolations similar to diatom-atom interactions.

The contour plots for the QSS dissociation reaction rate coefficient, rotational and vibrational energy transfer coefficients are presented in Figures \ref{fig:QSS_contour},\ref{fig:Figure2_Rot} and \ref{fig:Figure3_Vib}, respectively, for the $\mathrm{N_{2}+N}$ and $\mathrm{N_{2}+N_{2}}$ chemical systems at a translational temperature of 17500 K. Similar qualitative trends are observed across different translational temperatures. A visual examination of the contour plots provides useful insight into the temperature dependence of the respective coefficients and determination of generalized trends amongst different chemical systems. 

The behavior of the QSS dissociation rate coefficients for $\mathrm{N_{2}+N}$ system, as shown in Figure \ref{fig:QSS_N3}, represents a strong dependence on $\mathrm{T^{V}}$ for expansion flows ($\mathrm{T^{V}>>T^{R}}$) while a stronger dependence on $\mathrm{T^{R}}$ is observed for compression flows ($\mathrm{T^{R}>>T^{V}}$). Since the dissociation process is predominantly vibration-specific, these trends can be explained by considering the contribution to dissociation from vibrational levels in the two kinds of flows in strong non-equilibrium \cite{DaSilva}. For instance, when vibrational temperatures are low ($\mathrm{T^{R}>>T^{V}}$),  majority of the contribution comes from the low-lying vibrational levels due to their higher population, regardless of vibrational temperature. Therefore, the dissociation rate coefficients have a very weak dependence on vibrational temperature and are strongly correlated to the rotational temperature. On the other hand, at higher vibrational temperatures, the high-lying vibrational levels are significantly populated. Therefore, the contribution from high-lying vibrational levels is non-negligible and is highly dependent on the vibrational temperature. This results in a strong vibrational temperature dependence for cases with $\mathrm{T^{V}>>T^{R}}$. A contour plot for the diatom-diatom system ($\mathrm{N_{2}+N_{2}}$) is shown in Figure \ref{fig:QSS_N4}. The plot shows an overall qualitative trend similar to that observed in the diatom-atom system, with the rate coefficients now exhibiting appreciable dependence on both rotational and vibrational temperatures for compression flows. Under expansion-flow conditions, the dissociation rate coefficients remain strongly dependent on vibrational temperature, although this dependence is somewhat weaker than that observed for the $\mathrm{N_{2} + N}$ system. This behavior may also be partly attributed to some information lost at low rotational and vibrational temperatures due to quadratic extrapolation.

For the special case of a single bin, the net energy transfer coefficient is obtained by taking the difference between the forward and reverse energy transfer coefficients of the respective internal energy modes. Figures \ref{fig:Figure2_Rot} and \ref{fig:Figure3_Vib} show contour plots for the rotational and vibrational energy transfer coefficients, respectively, for both diatom-atom and diatom-diatom systems. For the sake of brevity and due to similar trends observed in oxygen systems, only the results for the nitrogen system are presented. A qualitative comparison between Figures \ref{fig:Rot_coeff_N3} and \ref{fig:Rot_coeff_N4} shows that, in general, the rotational energy transfer coefficients have a very strong dependence on the rotational temperature. This dependence weakens slightly at high vibrational temperatures over the entire range of rotational temperatures for $\mathrm{N_{2}+N}$, while the diatom-diatom system begins to show a weaker dependence on rotational temperature at high rotational and vibrational temperatures. Contour plots for the net vibrational energy transfer coefficients are shown in Figures \ref{fig:Vib_coeff_N3} and \ref{fig:Vib_coeff_N4} with the results showing similar trends between the two chemical systems. The vibrational energy transfer coefficients do not show as strong a dependence on its respective vibrational temperature as was observed in the case of rotational mode. At high vibrational and low rotational temperatures, the vibrational energy transfer coefficients show a stronger dependence on vibrational temperature which becomes weak as vibrational temperature goes down and rotational temperature increases. A strong dependence of rotational (vibrational) energy transfer coefficient on rotational (vibrational) temperature at high rotational (vibrational) temperatures is intuitive as the energy transfer contribution to the respective modes increases with the increase in the respective internal temperature due to higher population of the high-lying energy levels of the respective modes. Also, as mentioned before, since the macroscopic energy transfer coefficients are obtained by model reduction, these results are subject to information lost in reduction and curve-fitting, especially for the diatom-diatom system. 
 
\begin{figure}[h]
    \centering
    \includegraphics[width=0.95\linewidth]{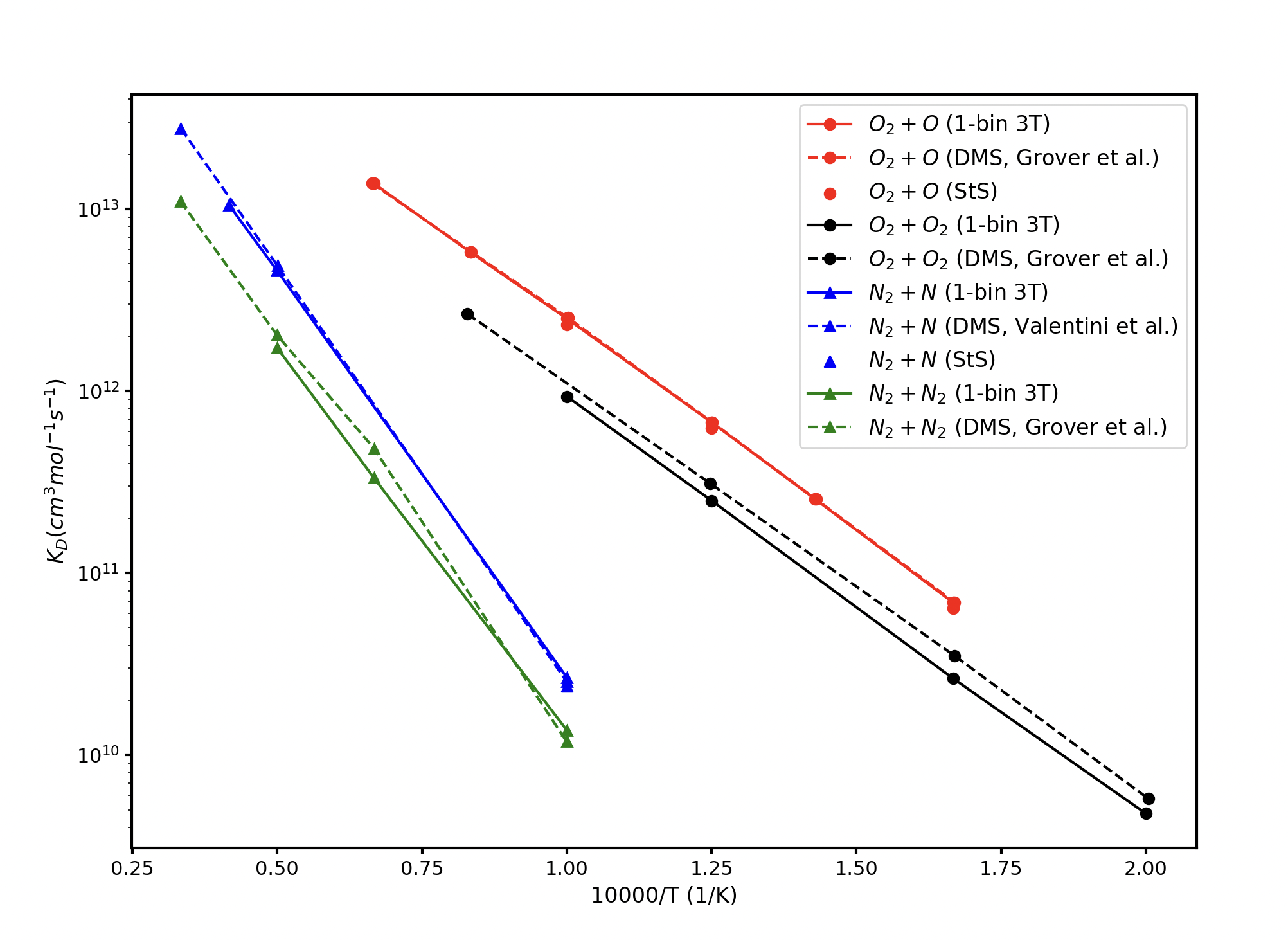}
    \caption{\textbf{Comparison of QSS dissociation rate constants with DMS \cite{Grover_2019_oxygen,Grover_N3N4_2022,Valentini_nitrogen_2016}}}
    \label{fig:QSS_KD}
\end{figure}

Figure \ref{fig:QSS_KD} shows a comparison of the QSS dissociation rate coefficients obtained using the 1-bin 3T model with those obtained by carrying out DMS simulations \cite{Grover_2019_oxygen,Grover_N3N4_2022,Valentini_nitrogen_2016} for all the chemical systems under consideration and a comparison with StS results is also shown for diatom-atom interactions. For the values pertaining to the 1-bin 3T model, the QSS dissociation rate coefficients were obtained using Eqs. \ref{Diss_rate_coeff} and \ref{Diss_ratecoeff_diatom-diatom} for diatom-atom and diatom-diatom systems, respectively, where the rotational and vibrational temperatures were determined from 0-D isothermal simulations. The results show a good agreement between the rate coefficients obtained from the model reduction and high-fidelity simulations, especially for the diatom-atom systems. A comparatively higher discrepancy observed in the diatom-diatom systems may be attributed to the unavailability of ro-vibrational StS kinetics data, which limits the ability to determine a more accurate non-equilibrium factor to model QSS distribution. Despite this, the non-equilibrium factor derived from DMS-obtained vibrational energy distribution data for diatom-diatom systems gives a reasonably good prediction of QSS dissociation rate coefficient.

\subsection{Vibrational and rotational relaxation time constants obtained using the 1-bin 3T model reduction}\label{sec:relax}
\begin{figure}[h]
    \centering
    \includegraphics[width=1\linewidth]{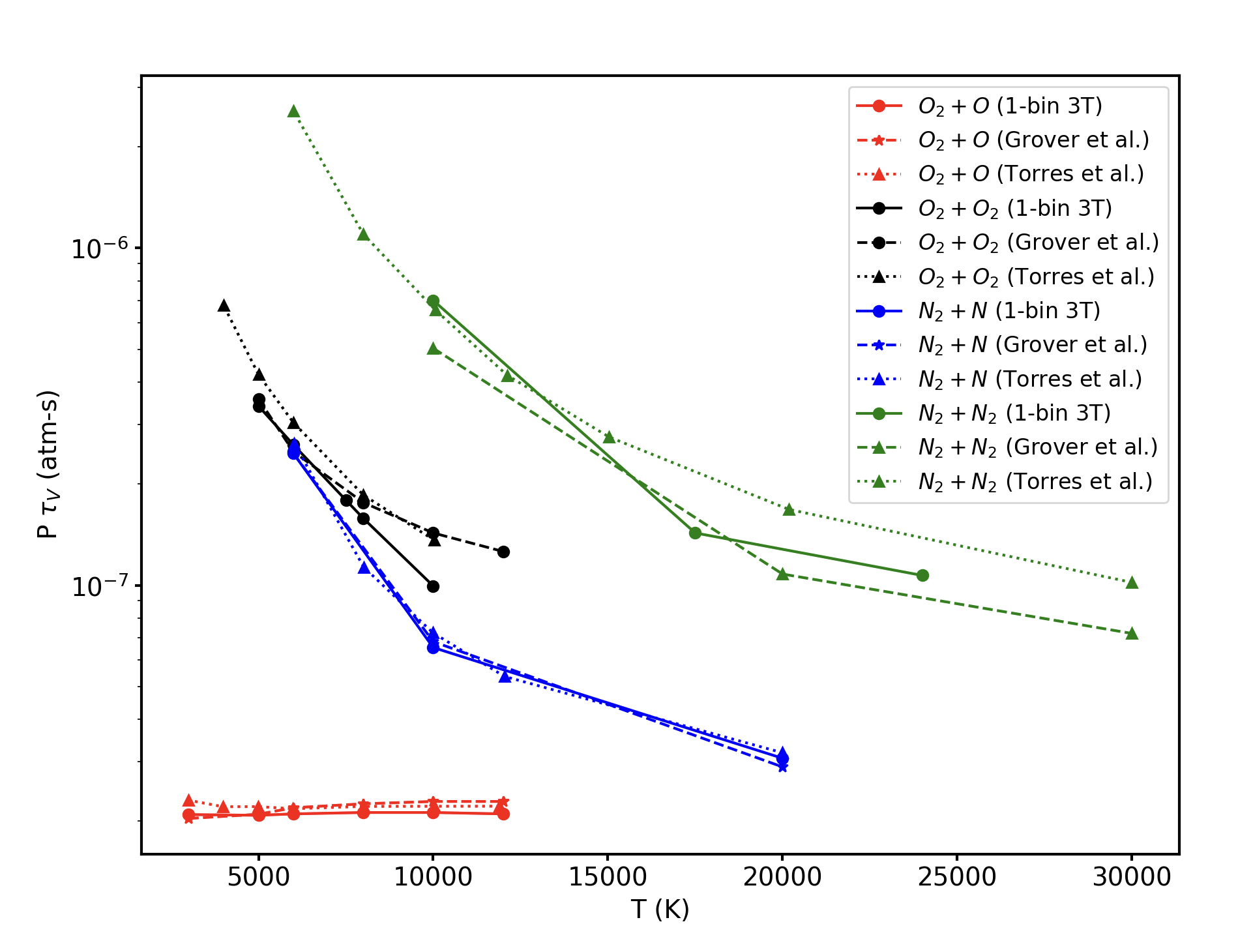}
    \caption{\textbf{Comparison of vibrational relaxation time constant with DMS \cite{Grover_2019_oxygen,Grover_N3N4_2022,Torres_DMS_Relaxation}}}
    \label{fig:Vib_relax}
\end{figure}

0-D isothermal simulations of a chemical system at a low initial internal temperature of 300 K subjected to different high heat bath temperatures were also carried out to determine the vibrational and rotational relaxation time constants of each chemical system using the e-folding method \cite{Vincenti-Kruger}. In order to study the internal energy relaxation phenomenon without the influence of chemical reactions, dissociation/recombination reactions were not included. Figure \ref{fig:Vib_relax} shows a comparison of the vibrational relaxation time constants obtained using the 1-bin 3T model with those obtained from DMS simulations \cite{Grover_2019_oxygen,Grover_N3N4_2022,Torres_DMS_Relaxation}. In general, there is a good agreement between the time constants obtained using the two methods with a comparatively higher discrepancy observed in the case of diatom-diatom systems. It must be noted that there is also a discrepancy observed in DMS values obtained from different sources. This discrepancy can stem from different time steps used with the Velocity Verlet time integration scheme, impact parameter and initial distance between collision partners \cite{Torres_DMS_Relaxation}. Uncertainty can also stem from statistical noise due to finite sample sizes \cite{Torres_DMS_Relaxation}. The vibrational relaxation time constants obtained using the 1-bin 3T model for $\mathrm{O_{2}+O_{2}}$ system have a 5 \% error at low temperatures when compared to DMS simulations by Grover et al.\cite{Grover_2019_oxygen} However, this error increases to 30 \% at 10000 K. In contrast, the results show an error close to 20 \% when compared against DMS results from Torres et al.\cite{Torres_DMS_Relaxation} at all translational temperatures except 10000 K, where the error is same as that observed against the DMS calculations of Grover et al. On the contrary, the 1-bin 3T results for $\mathrm{N_{2}+N_{2}}$ show a maximum error of $\approx$ 38 \% with DMS results from Grover et al.\cite{Grover_N3N4_2022} at T=10000 K and an error of 6.87 \% at the same translational temperature when compared against results obtained from Torres et al.\cite{Torres_DMS_Relaxation}. At higher translational temperatures, the 1-bin 3T model gives values closer to the predictions by Grover et al. Despite the use of curve fitting at low internal temperatures for diatom-diatom systems, the results obtained from the 1-bin 3T model are able to capture the vibrational relaxation phenomenon quantitatively and qualitatively well for all the chemical systems considered. 

\begin{figure}[h]
    \centering
    \includegraphics[width=1\linewidth]{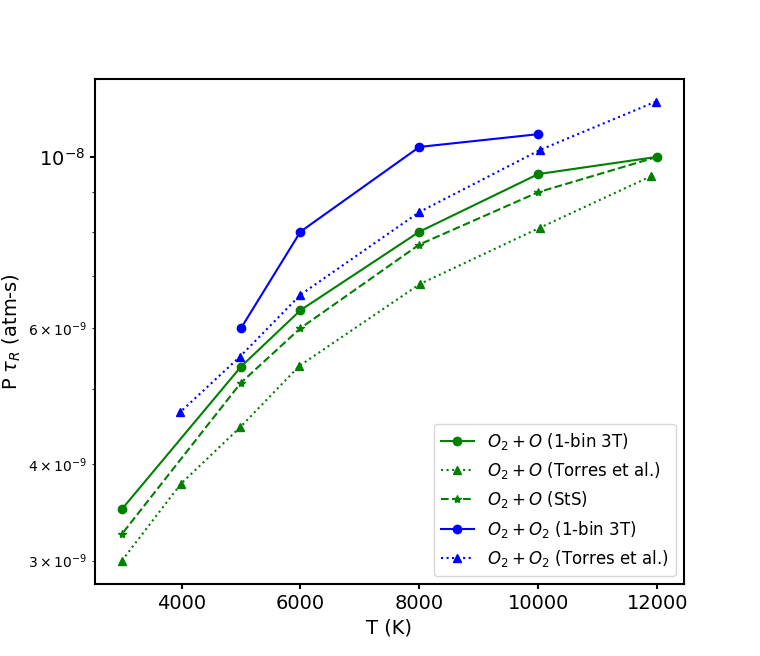}
    \caption{\textbf{Comparison of rotational relaxation time constant for $O_{2}+O$ and $O_{2}+O_{2}$ collisions with DMS \cite{Torres_DMS_Relaxation}}}
    \label{fig:Rot_oxygen}
\end{figure}

\begin{figure}[h]
    \centering
    \includegraphics[width=1\linewidth]{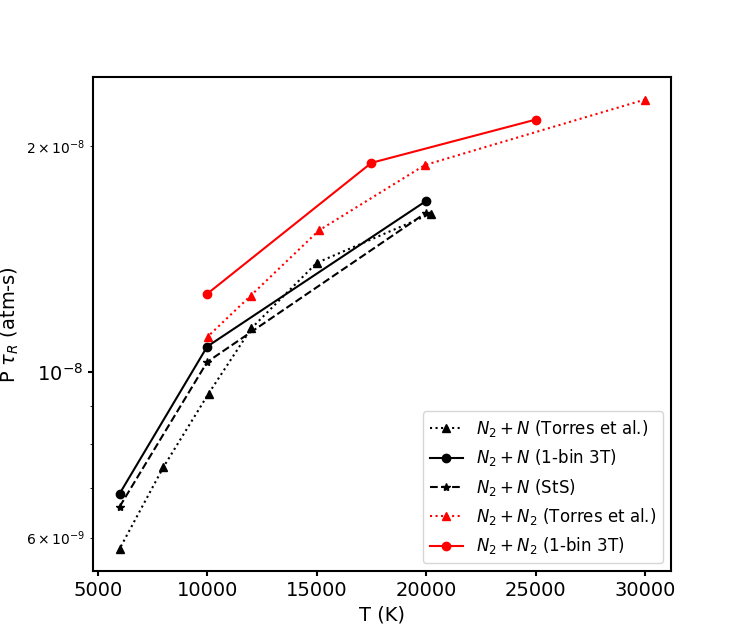}
    \caption{\textbf{Comparison of rotational relaxation time constant for $N_{2}+N$ and $N_{2}+N_{2}$ collisions with DMS\cite{Torres_DMS_Relaxation}}}
    \label{fig:Rot_nitrogen}
\end{figure}

Rotational relaxation phenomenon was also analysed by obtaining the rotational relaxation time constants using the e-folding method, as shown in Figures \ref{fig:Rot_oxygen} and \ref{fig:Rot_nitrogen} for oxygen and nitrogen systems, respectively. Ro-vibrational StS analysis was carried out for diatom-atom interactions in addition to DMS results obtained from literature for comparison with the 1-bin 3T model. The results obtained from the 1-bin 3T model show good qualitative agreement with the higher fidelity solutions. Compared with the DMS results \cite{Torres_DMS_Relaxation}, the 1-bin 3T model overpredicts the rotational relaxation time of the $\mathrm{O_{2}+O}$ system by about 15--20\% between 3000 and 10000 K, and by about 6\% at 12000 K. However, the results show a better agreement with the StS results, having a lower error of $\approx$ 5 \%. The rotational relaxation time constants obtained from the 1-bin 3T model for $\mathrm{O_{2}+O_{2}}$ chemical system also show a discrepancy with the DMS results\cite{Torres_DMS_Relaxation} and are off by a factor of $\approx$ 1.2 at 6000 K and 8000 K. Similar to the oxygen system, diatom-atom interactions involving nitrogen molecule also show a better agreement with StS results as compared to DMS results\cite{Torres_DMS_Relaxation} and the model obtained value is off by a factor of 1.15 at 10000 K for diatom-diatom interactions, as shown in Figure \ref{fig:Rot_nitrogen}. The validation analysis of rotational relaxation time constants highlights the ability of the 1-bin 3T model to satisfactorily capture the rotational relaxation phenomenon, which is generally assumed to be in equilibrium with the translational mode due to its fast relaxation behavior, in widely used multi-temperature models.

\begin{figure}[h]
    \centering
    \includegraphics[width=1\linewidth]{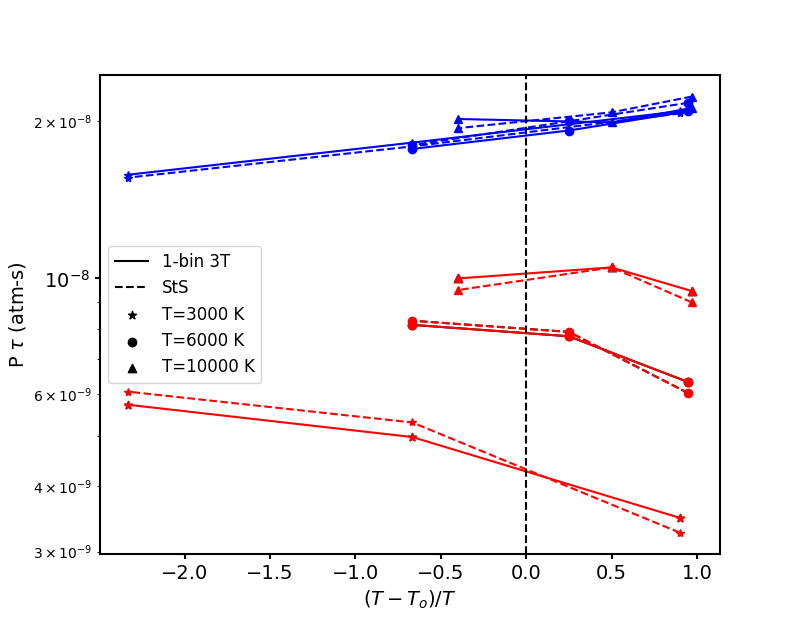}
    \caption{\textbf{Relaxation time constants in heating vs cooling scenarios for $O_{2}+O$ system. Red and blue colors represent rotational and vibrational relaxation, respectively.}}
    \label{fig:O3_Relax}
\end{figure}

\begin{figure}[h]
    \centering
    \includegraphics[width=1\linewidth]{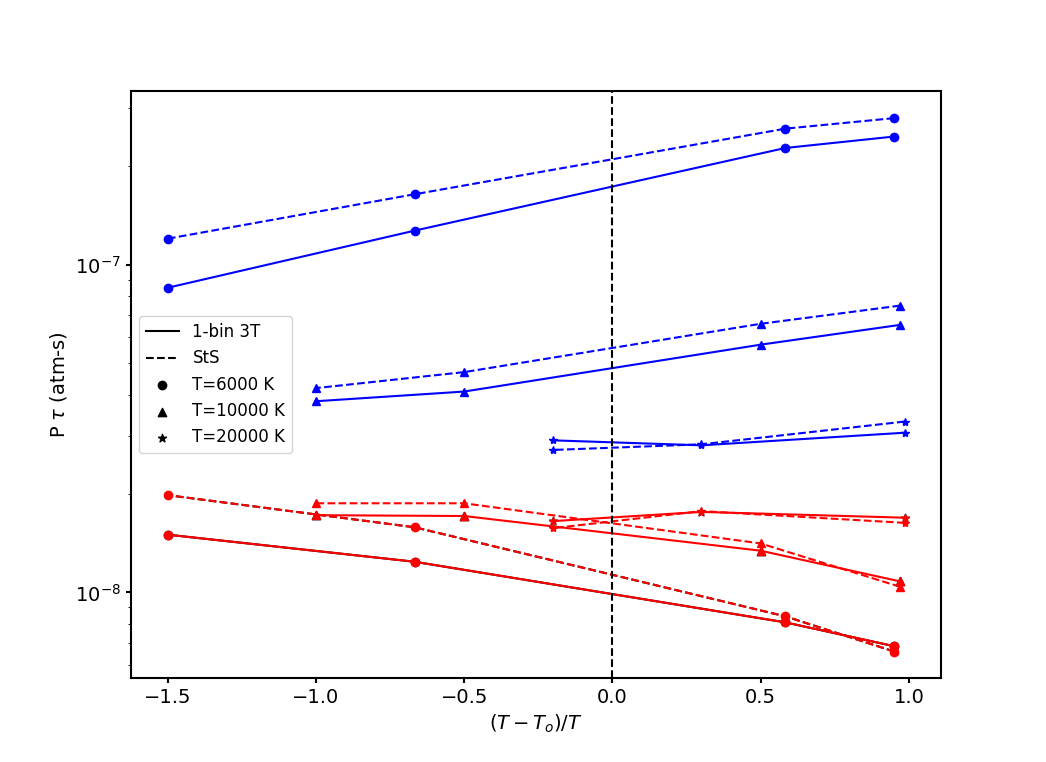}
    \caption{\textbf{Relaxation time constants in heating vs cooling scenarios for $N_{2}+N$ system. Red and blue colors represent rotational and vibrational relaxation, respectively.}}
    \label{fig:N3_Relax}
\end{figure}

Up to this point, all cases analysed in this section pertained to a heating/dissociating environment in which the chemical system initially at a low internal temperature is suddenly subjected to high heat bath temperatures to induce thermo-chemical non-equilibrium. While this scenario is mainly a characteristic of shocks in hypersonic environments, an inverse situation of expanding flows is predominant around shoulder of the vehicle and in nozzles. Previous high-fidelity StS \cite{Notey-Recomb-JCP2025} and DMS studies\cite{Torres_DMS_Relaxation} have shown that the relaxation behavior is different for expanding and compressing flows. This is because the relaxation behavior is highly dependent on the initial distribution of internal energy levels, which is dictated by the initial conditions. 0-D simulations were carried out at different initial internal temperature conditions for multiple heat bath temperatures to validate the ability of the 1-bin 3T model to predict the relaxation behavior for both expanding and compressing flows, the results of which are shown in Figures \ref{fig:O3_Relax} and \ref{fig:N3_Relax}. In general, the results obtained from the 1-bin 3T model show a good agreement with StS results for both nitrogen and oxygen systems. A visual examination of the results shows a lower error incurred for the $\mathrm{O_{2}+O}$ system when compared to the $\mathrm{N_{2}+N}$ system. The 1-bin 3T model reduction in $\mathrm{N_{2}+N}$ case shows a larger discrepancy with the StS results for expanding cases and strong non-equilibrium conditions. A plausible explanation for this behavior could be that the relaxation dynamics of internal energy levels for $N_{2}+N$ system is characterized by an instantaneous population distribution which cannot be exactly represented by the Maxwell-Boltzmann distribution (Eq.~(\ref{Pop_distribution_function})) at instantaneous rotational and vibrational temperatures due to a larger energy exchange barrier in the low-lying energy levels as compared to the oxygen system. Nevertheless, the maximum discrepancy incurred by the $\mathrm{N_{2}+N}$ system in the cases considered is a factor of 0.75 at T=6000 K with an initial internal temperature $T_{o}$=15000 K. It is also interesting to point out that Notey et al.\cite{Notey-Recomb-JCP2025} observed that the relaxation dynamics of $\mathrm{N_{2}+N}$ system at low heat bath temperatures deviates from Landau-Teller relaxation \cite{Landau-Teller} when subjected to very strong non-equilibrium while Landau-Teller relaxation accurately captures the dynamics for $\mathrm{O_{2}+O}$ system.

\subsection{Application of the 1-bin 3T model to diatom-atom interactions 
$(\mathrm{O_{2}+O}$ and $\mathrm{N_{2}+N})$ in a 0-D isothermal reactor}
\begin{figure}[h]
    \centering
    \includegraphics[width=1\linewidth]{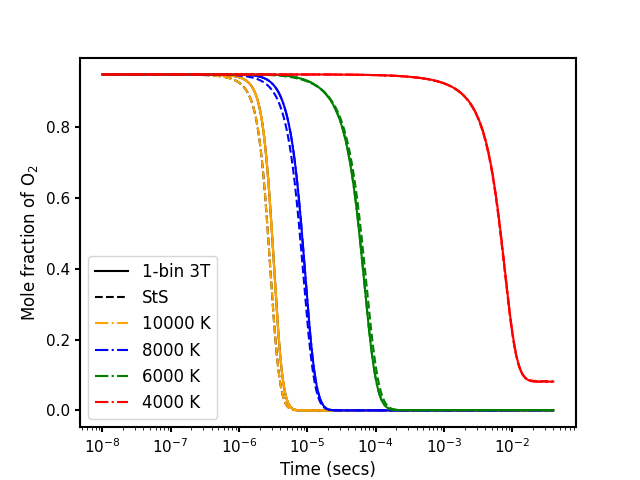}
    \caption{\textbf{Comparisons of the mole fraction profiles of $\mathrm{O_{2}}$ in a 0-D isothermal reactor at different heat bath temperatures for $\mathrm{O_{2}+O}$ chemical system.}}
    \label{fig: O3 isothermal Mf}
\end{figure}

\begin{figure}[]
\centering
    \subfigure[]
    {\label{fig:s1}\includegraphics[width=0.45\textwidth]{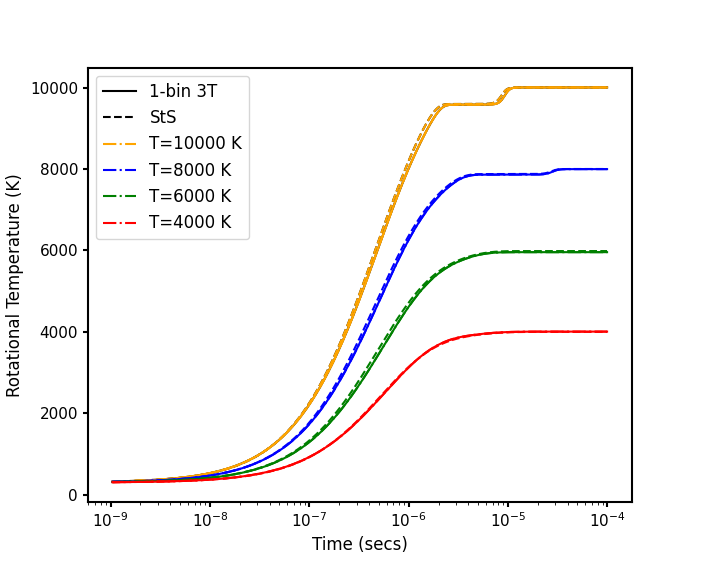}}
    \subfigure[]{\label{fig:s2}\includegraphics[width=0.47\textwidth]{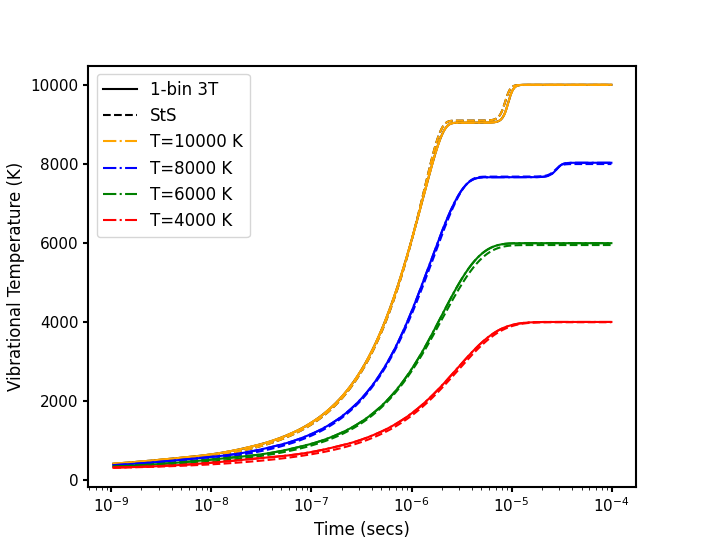}}
    \caption{\textbf{Comparison of (a) Rotational and (b) Vibrational temperature evolution in a 0-D isothermal reactor at different heat bath temperatures for $\mathrm{O_{2}+O}$ chemical system}}
    \label{fig: 0-D iso O3}
\end{figure}

This section aims to validate the 1-bin 3T model against StS 0-D isothermal simulations for $\mathrm{O_{2}+O}$ and $\mathrm{N_{2}+N}$ chemical systems in both dissociating and recombining environments by comparing the mole fraction and internal temperature evolution plots obtained from the 1-bin 3T model against those obtained from StS simulations. The 0-D isothermal reactor governing equations for diatom-atom collisions are given by Eqs. \ref{Diatom-atom-iso1}, \ref{Diatom-atom-iso2} and \ref{Diatom-atom-iso3}, where the rate of change of mass and internal energies of the system is equated to the respective kinetics source terms on the right side of the equations. This analysis also aims to confirm the applicability of a single non-equilibrium factor across a wide range of temperature conditions and identify the conditions where its accuracy might be compromised.

\subsubsection{Dissociating environment}
The 1-bin 3T model was applied to run 0-D isothermal simulations for a wide range of translational temperature conditions from 4000 K to 10000 K for $\mathrm{O_{2}+O}$ chemical system.
 The system initialized at a low internal temperature of 300 K and an initial pressure of 1000 Pa is suddenly heated to a high heat bath temperature to simulate non-equilibrium dissociation. The mixture initially consists of 95\% oxygen molecules and 5\% oxygen atoms by moles.

The mole fraction evolution for the different heat bath temperatures considered is shown in Figure \ref{fig: O3 isothermal Mf}. The results show a good overall agreement between model reduction and the high-fidelity StS solution for all heat bath temperatures considered. This supports the applicability of a single functional form of the non-equilibrium factor to a range of temperature conditions. The rotational and vibrational temperature evolution for the different heat bath cases is shown in Figures \ref{fig: 0-D iso O3} (a) and \ref{fig: 0-D iso O3}(b), respectively, with the results showing an accurate capture of the rotational and vibrational energy transfer phenomenon by the proposed 1-bin 3T model. 

\begin{figure}[h]
    \centering
    \includegraphics[width=1\linewidth]{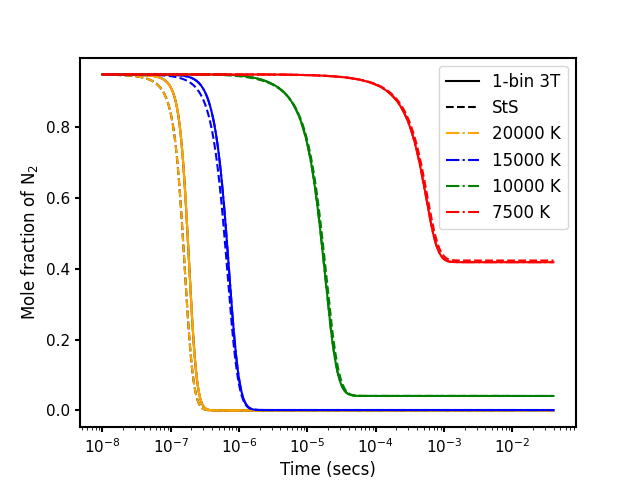}
    \caption{\textbf{Comparisons of the mole fraction profiles of $\mathrm{N_{2}}$ in a 0-D isothermal reactor at different heat bath temperatures for $\mathrm{N_{2}+N}$ chemical system.}}
    \label{fig:Mf_iso_N3}
\end{figure}

\begin{figure*}
\centering
    \subfigure[]{\label{fig:s1}\includegraphics[width=0.45\textwidth]{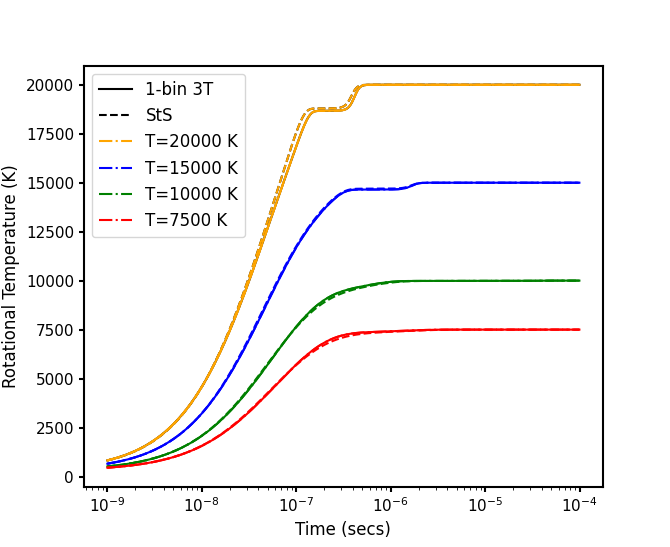}}
    \subfigure[]{\label{fig:s2}\includegraphics[width=0.47\textwidth]{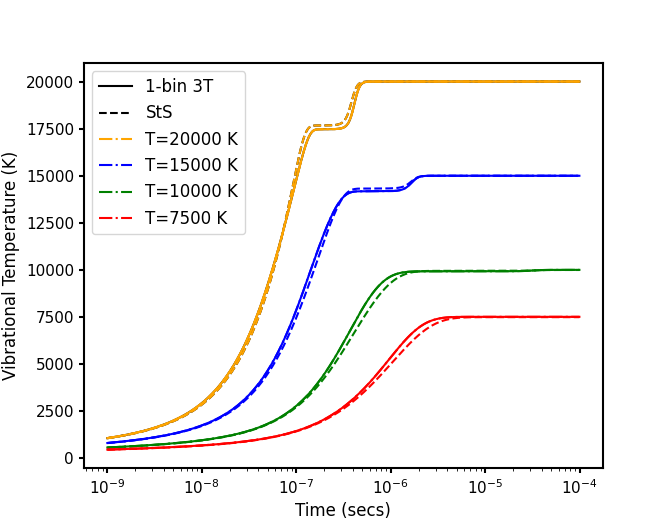}}
    \caption{\textbf{Comparison of (a) Rotational and (b) Vibrational temperature evolution in a 0-D isothermal reactor at different heat bath temperatures for $\mathrm{N_{2}+N}$ chemical system}}
    \label{fig: Temp_iso_N3}
\end{figure*}

While the high heat bath temperatures are characterized by almost complete dissociation at equilibrium, the lowest heat bath temperature of 4000 K establishes an equilibrium stage at final molecular mole fraction of around 0.081, as shown in Figure \ref{fig: O3 isothermal Mf}. This is a result of recombination processes which are not entirely insignificant at low heat bath temperatures. Dissociation and recombination processes balance each other to establish an early equilibrium stage. The results show that the 1-bin 3T model is able to accurately capture this dissociation/recombination phenomenon at low temperatures. As the translational temperature increases, it is observed that the disparity between the 1-bin 3T model and StS simulation increases. For instance, the mole fraction evolution obtained from the 1-bin 3T model at T=10000 K shows some discrepancy with StS during the start of bulk dissociation (\emph{e.g.}, $t \simeq 10^{-6}$s). A plausible explanation for this is a coupling between energy transfer and dissociation processes, which the 1-bin 3T model in its proposed form is not formulated to capture. The rotational and vibrational temperature evolution plots show that the energy transfer phenomenon is not completed by the start of bulk dissociation for temperatures higher than  6000 K. A coupling with chemistry influences the energy transfer dynamics which results in a slight difference in the evolution plots at high temperatures. While the 1-bin 3T model works well for translational temperatures up to 10000 K, higher temperatures would require an additional QSS modeling for energy transfer processes to accurately capture the dynamics. Additionally, the 1-bin 3T model provides a substantial computational speedup over the StS model, requiring only 0.561 s of CPU time as compared to 610.61 s for the corresponding isothermal StS simulation at 10,000 K on a single CPU core— a three-orders-of-magnitude reduction in computational cost.

The 1-bin 3T model was validated by performing 0-D isothermal reactor simulations over translational temperatures ranging from 7500-20000 K for the $\mathrm{N_{2}+N}$ chemical system as well. The system was initialized at a low internal temperature of 300 K and an initial pressure of 10000 Pa and suddenly heated to their respective high heat bath temperatures to simulate non-equilibrium dissociation. The mixture initially consists of 95\% nitrogen molecules and 5\% nitrogen atoms by moles. An observation similar to the mole fraction evolution of $\mathrm{O_{2}+O}$ system was made for the $\mathrm{N_{2}+N}$ system, as shown in Figure \ref{fig:Mf_iso_N3}. When considering the internal temperature evolution, Figures \ref{fig: Temp_iso_N3} (a) and (b) show that while the 1-bin 3T model is able to capture the dynamics reasonably well for rotational mode, there is a visible discrepancy observed in the evolution of vibrational temperature as the system approaches equilibrium temperature. As already discussed in Sec.~\ref{sec:relax}, this discrepancy arises due to relaxation dynamics and population distribution of nitrogen not being exactly captured by Eq.~(\ref{Pop_distribution_function}). However, similar to $\mathrm{O_{2}+O}$, the model reduction is able to predict the QSS plateau in temperature evolution at high translational temperatures for $\mathrm{N_{2}+N}$ as well.

\begin{figure*}[hbt!]
\centering
    \subfigure[]{\label{fig:s1}\includegraphics[width=0.45\textwidth]{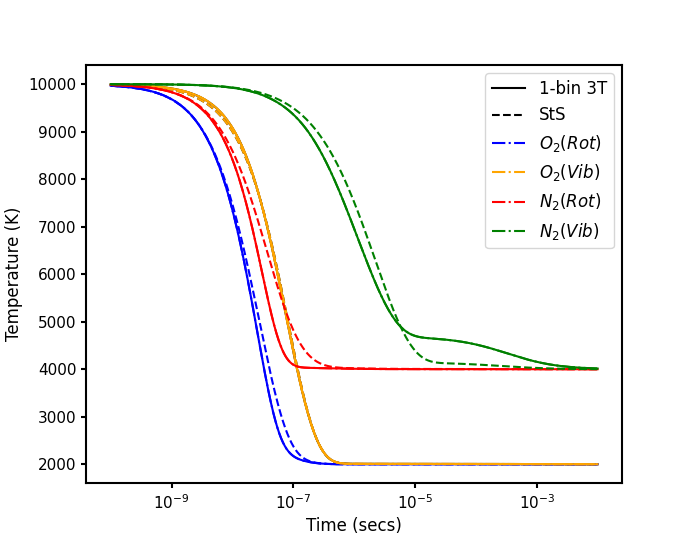}}
    \subfigure[]{\label{fig:s2}\includegraphics[width=0.47\textwidth]{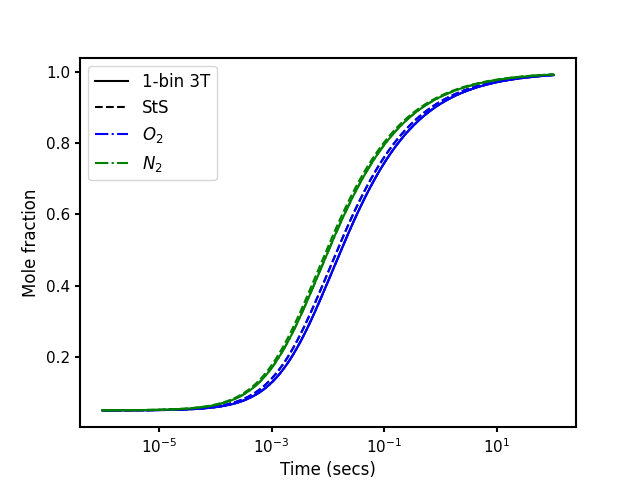}}
    \caption{\textbf{Comparisons of (a) temperature and (b) molecular species mole fraction evolution in a recombining 0-D isothermal reactor for $\mathrm{O_{2}+O}$ and $\mathrm{N_{2}+N}$ chemical system.}}
    \label{fig: 0-D recomb}
\end{figure*}

\subsubsection{Recombining environment}
\begin{figure*}[hbt!]
\centering
    \subfigure[]{\label{fig:s1}\includegraphics[width=0.45\textwidth]{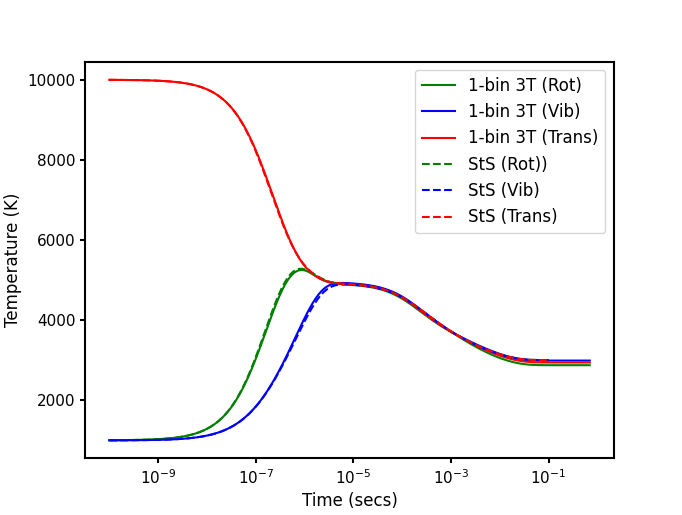}}
    \subfigure[]{\label{fig:s2}\includegraphics[width=0.45\textwidth]{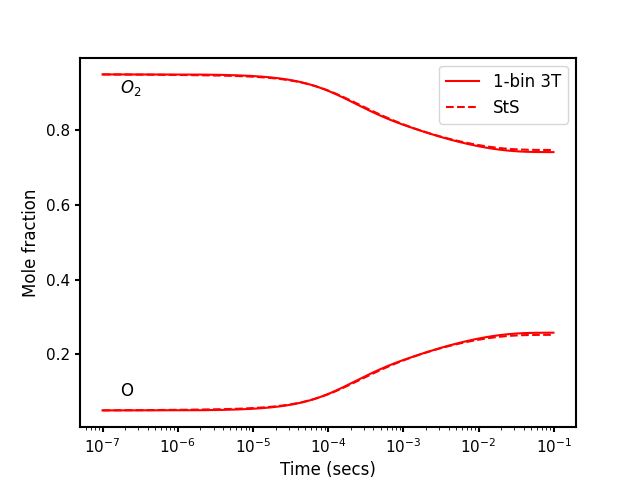}}
    \caption{\textbf{Comparisons of (a) temperature and (b) species mole fraction evolution in the 0-D adiabatic for $O_{2}+O$ chemical system.}}
    \label{fig: 0-D adiabatic O3}
\end{figure*}

\begin{figure*}[hbt!]
\centering
    \subfigure[]{\label{fig:s1}\includegraphics[width=0.45\textwidth]{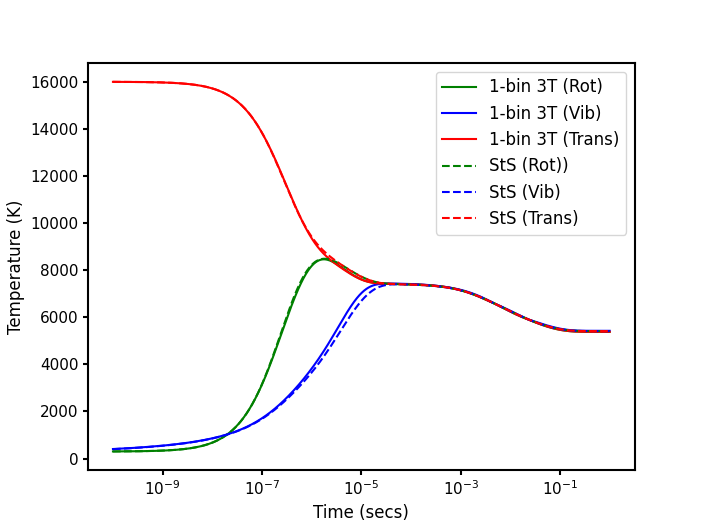}}
    \subfigure[]{\label{fig:s2}\includegraphics[width=0.45\textwidth]{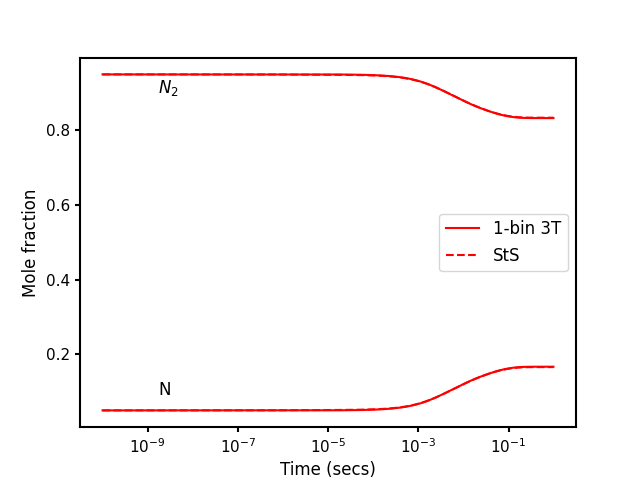}}
    \caption{\textbf{Comparisons of (a) temperature and (b) species mole fraction evolution in the 0-D adiabatic reactor for $N_{2}+N$ chemical system.}}
    \label{fig:N3_adiabatic}
\end{figure*}

While dissociation is the dominant process within shock layers, recombination becomes prevalent in the expanding regions of the flow, such as downstream of blunt-body shoulders and within wake regions. The relaxation and reactive dynamics in recombining flows differ fundamentally from those in dissociating environments \cite{Notey-Recomb-JCP2025, Panesi2013_N3}. In particular, recombination leads to a significantly larger population of high-lying vibrational states compared to dissociating flows. Owing to the strong coupling between chemistry and vibrational energy modes, recombination adds energy into the vibrational mode, thereby sustaining elevated vibrational populations and slowing equilibration of the vibrational temperature. The increased population of high-energy vibrational states also amplifies anharmonic effects, which are less pronounced in dissociating environments due to the preferential depletion of highly excited vibrational levels during dissociation \cite{Notey-Recomb-JCP2025}. Because the non-equilibrium correction factor employed in the present model reduction framework is derived from quasi-steady-state (QSS) distributions characteristic of dissociating environments, it is essential to assess the accuracy and robustness of the 1-bin 3T model under recombining conditions, where the underlying QSS behavior differs substantially.

To that end, 0-D simulations of $\mathrm{O_{2}+O}$ and $\mathrm{N_{2}+N}$ systems under isothermal conditions were carried out at fixed low heat bath temperatures of 2000 K and 4000 K, respectively, to simulate environments dominated by recombination. Both the systems were initialized at a high internal temperature of 10000 K and pressure of $\mathrm{10^{5}}$ Pa, and suddenly cooled to the corresponding low-heat bath temperatures. The initial mole fraction of atomic species is taken to be 0.95 and that of molecular species is 0.05.

Figure \ref{fig: 0-D recomb} (a) shows the rotational and vibrational temperature evolution obtained from StS simulations and the 1-bin 3T model reduction for both $\mathrm{N_{2}+N}$ and $\mathrm{O_{2}+O}$ chemical systems. The mole fraction evolution of the molecular species is shown in Figure \ref{fig: 0-D recomb} (b). The StS results show that for the recombining conditions considered, the bulk recombination occurs after the chemical systems have achieved thermal equilibrium. While the 1-bin 3T model captures the temperature evolution reasonably well for the $\mathrm{O_{2}+O}$ system, the model fails to predict the temperature evolution of $\mathrm{N_{2}+N}$ system to the same accuracy. The rotational temperature relaxation predicted by the 1-bin 3T model for $\mathrm{N_{2}+N}$ is faster as compared to the StS results. A plausible cause of this is an inadequate representation of the StS population distribution in a cooling environment by Eq. (5) due to different rotational temperatures pertaining to each vibrational state, with the quasi-bound states exhibiting higher rotational temperature \cite{Notey-Recomb-JCP2025}. These complex distributions are not accurately captured by a simplified representation using a single rotational and single vibrational temperature. The vibrational temperature of $\mathrm{N_{2}}$ predicted by the 1-bin 3T model stops decreasing at about $t=10^{-5}$ s and remains nearly constant until about $10^{-4}$ s, which corresponds to the onset of recombination. Due to strong vibrational coupling of the recombination phenomenon, during the early stages, the 1-bin 3T model predicts a higher energy added to the vibrational mode as compared to StS results. Despite the differences observed in internal temperature evolution, the 1-bin 3T model predicts the mole fraction evolution to an excellent agreement with the StS simulations which highlights the ability of applying the non-equilibrium factor obtained from a dissociation dominated environment to recombining conditions. An analogous result was also observed by Notey et al.\cite{Notey-Recomb-JCP2025} in their study of recombination phenomenon where it was concluded that the value of the recombination rate constant calculated by applying detailed balance to the QSS dissociation rate constant (obtained by simulating a dissociating environment at the concerned heat bath temperature) was very close to that of an effective recombination rate constant predicted using ro-vibrational StS analysis of the recombining environment. 

\subsection{Application of the 1-bin 3T model to diatom-atom interactions 
$(\mathrm{O_{2}+O}$ and $\mathrm{N_{2}+N})$ in a 0-D adiabatic reactor}
\begin{figure*}
\centering
    \subfigure[]{\label{fig:s1}\includegraphics[width=0.45\textwidth]{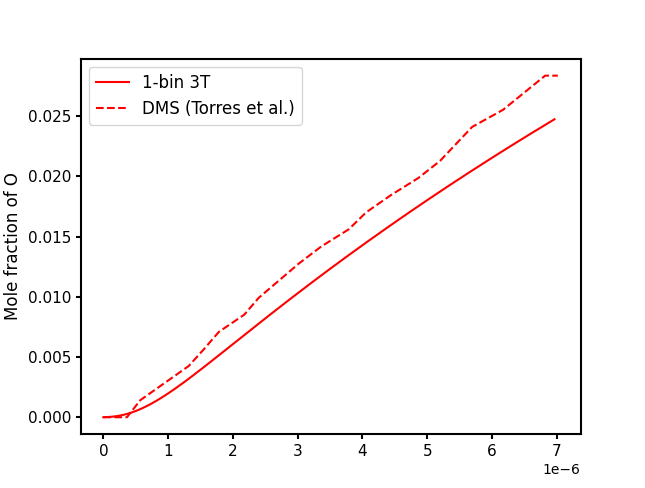}}
    \subfigure[]{\label{fig:s2}\includegraphics[width=0.44\textwidth]{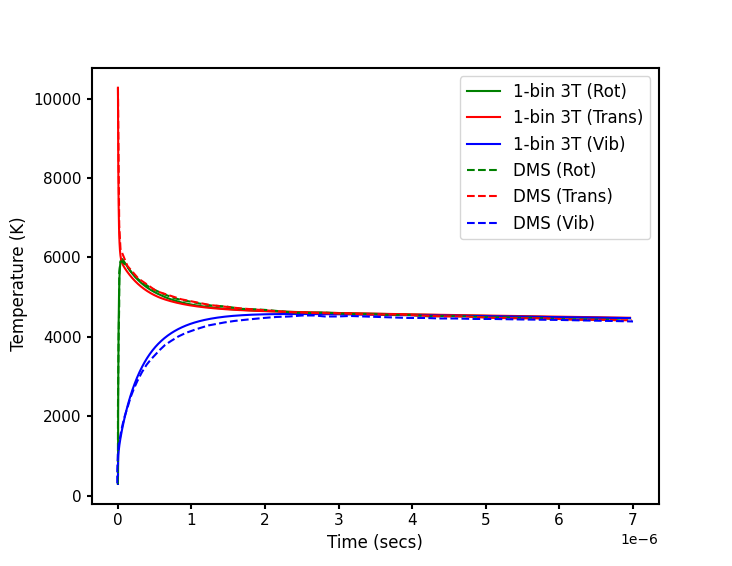}}
    \caption{\textbf{Comparison of (a) mole fraction and (b) temperature evolution in a 0-D adiabatic reactor for complete oxygen mixture with DMS (Torres et al.) \cite{Adiabatic_oxygen}}}
    \label{fig: Adiabatic}
\end{figure*}

The 1-bin 3T model was also tested in adiabatic conditions to check its validity in a non-equilibrium system over a range of evolving translational temperature conditions. A 0-D adiabatic reactor case is characterized by a change in the translational temperature, in addition to the number density and internal energy modes. An adiabatic simulation setup results in a transfer of energy between translational and internal energy modes, with the translational energy also affecting chemical energy, thereby influencing the dissociation process. The system of equations for the 0-D adiabatic reactor must be supplemented by a governing equation describing the evolution of the total energy, as described in detail by Munaf\`o et al.\cite{Munafo_JTHT_plato_2025}


Adiabatic simulation involving $\mathrm{O_{2}+O}$ chemical system at an initial internal temperature of 1000 K and an initial translational temperature of 10000 K was performed with 95\% oxygen molecules and 5\% oxygen atoms at an initial pressure of 5000 Pa. Similarly, a 0-D adiabatic simulation involving $\mathrm{N_{2}+N}$ chemical system at an initial internal temperature of 300 K and translational temperature of 16000 K was performed with 95\% nitrogen molecules and 5\% nitrogen atoms at an initial pressure of 1000 Pa. These test cases validate the accuracy of the 1-bin 3T model in a predominantly dissociating environment with a non-negligible recombination rate. 

The translational, rotational and vibrational temperature evolution profiles for $\mathrm{O_{2}+O}$ are shown in Figure \ref{fig: 0-D adiabatic O3} (a). The results obtained using the 1-bin 3T model are in excellent agreement with the StS predictions. The figure shows that the translational and rotational temperatures quickly thermalize with each other around 1.5 $\mu$s while the vibrational temperature lags behind and approaches the equilibrated trans-rotational temperature around 4 $\mu$s. This is followed by a short period of temperature plateau after which the temperatures begin to drop due to dissociation.

The mole fraction evolution profiles for molecular and atomic oxygen are shown in Figure \ref{fig: 0-D adiabatic O3} (b). The dissociation process begins after the translational and internal modes have equilibrated with each other. The comparison with the StS results demonstrates that the 1-bin 3T model accurately captures the dissociation/recombination phenomenon both quantitatively and qualitatively. The error between the equilibrium mole fraction of oxygen atom predicted by 1-bin 3T model and StS is minor ($\approx$ 0.006). A plausible explanation for this error may be attributed to errors incurred during interpolation between translational temperatures when using group tables. As can be seen in Figure \ref{fig: 0-D adiabatic O3} (a), the equilibrium translational and internal temperatures appear to have a minor difference which results from unavailability of group tables data at that particular translational temperature and reliance on interpolations to obtain the required value of reaction rate and energy transfer coefficients. 

The adiabatic reactor evolves similarly for $\mathrm{N_{2}+N}$ chemical system as well, as shown in Figures \ref{fig:N3_adiabatic}(a) and (b). As observed in the isothermal reactor case, the vibrational temperature evolution exhibits minor discrepancies close to the peak vibrational temperature. However, post that stage, the 1-bin 3T model is able to accurately capture both the energy transfer and dissociation behavior. The temperature evolution results show the attainment of equilibrium temperature for all the modes, as opposed to slight discrepancies observed in $\mathrm{O_{2}+O}$ at equilibrium. Similarly, the equilibrium mole fraction of nitrogen atoms and molecules is also predicted accurately by the 1-bin 3T model. This is a result of a better interpolation in group tables which may be influenced by the number of points in the group table for both exact and curve-fitted data. An accurate prediction of the equilibrium mole fractions for both the chemical systems considered also confirms the robustness of the non-equilibrium factor in simulation environments that have competing dissociation and recombination processes.

\subsection{Application of the 1-bin 3T model to pure mixtures including both diatom-diatom and diatom-atom interactions in a 0-D adiabatic reactor}

In this section, 0-D adiabatic results for low enthalpy cases of pure nitrogen and oxygen mixtures involving both diatom-atom and diatom-diatom interactions are presented. 
Adiabatic simulations for pure molecular oxygen mixture of density 0.03446 $kg/m^{3}$ at an initial translational and internal temperatures of 10273 K and 298.2 K, respectively, are performed and validated against DMS \cite{Adiabatic_oxygen}. The results in Figure \ref{fig: Adiabatic}(a) show that the 1-bin 3T model underpredicts the O mole fraction throughout the simulated time, by about 30\% at 1 $\mu$s and about 13\% at 7 $\mu$s (absolute difference below $4\times10^{-3}$). This low-enthalpy case dissociates only weakly ($x_{O}<0.03$ at 7 $\mu$s) and the systematic underprediction is consistent with the $\mathrm{O_{2}+O_{2}}$ QSS dissociation rate coefficient being 15--25\% below DMS in Figure \ref{fig:QSS_KD}. The DMS case was simulated for 7 $\mu$s. By about 3 $\mu$s the translational, rotational and vibrational temperatures have relaxed to a common value of about 4500 K, while the composition is still far from chemical equilibrium. It must be noted that the dissociation rate coefficients for the simulated time mainly come from the extrapolated values at temperatures lower than 5000 K.  For the simulated time, the temperature evolution results obtained by 1-bin 3T model in  Figure \ref{fig: Adiabatic} (b) show a good agreement with the DMS results, thereby confirming the accuracy of the proposed 1-bin 3T model over the evolving translational temperature conditions while significantly reducing the computational cost. 

\begin{figure}[h]
    \centering
    \includegraphics[width=1\linewidth]{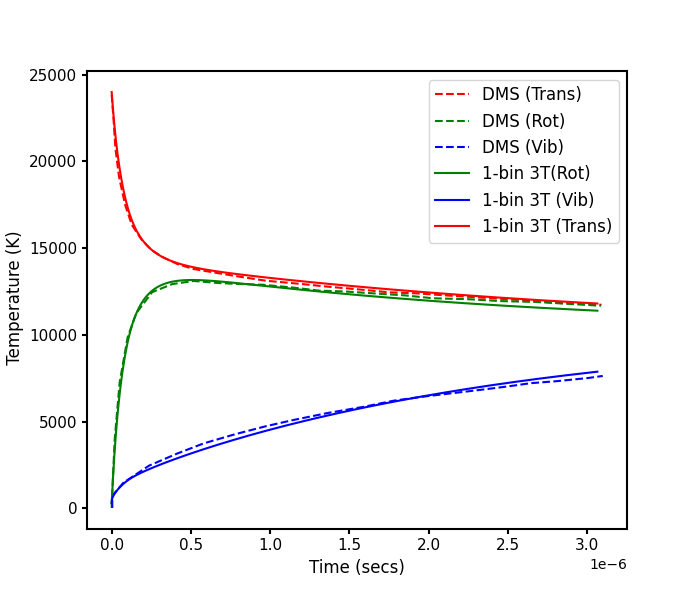}
    \caption{\textbf{Comparisons of the temperature profiles of $\mathrm{N_{2}}$ in a 0-D adiabatic reactor for pure nitrogen mixture with DMS (Torres et al.) \cite{Torres_adiabatic_N2_2020}}}
    \label{fig:N3N4_adia}
\end{figure}
A similar comparison for the pure nitrogen mixture with density 0.0021 $kg/m^{3}$ was carried out by simulating adiabatic conditions initialized at an internal and translational temperature of 300 K and 24000 K, respectively. Figure \ref{fig:N3N4_adia} shows the evolution plot of the temperatures of each mode over the simulated period of time (3 $\mu$s). As opposed to the low-enthalpy case considered for oxygen, the nitrogen system does not reach thermal equilibrium by the end of 3 $\mu$ s and undergoes negligible dissociation. Therefore, only the results for internal temperature evolution are presented. Figure \ref{fig:N3N4_adia} shows good agreement between the results obtained from the 1-bin 3T model and DMS \cite{Torres_adiabatic_N2_2020} with the results showing similar qualitative trend as the oxygen mixture.

\section{Summary and Conclusions} 
A three-temperature model involving a single bin was proposed to model the non-equilibrium behavior of high-temperature gas mixtures, and it was demonstrated for pure oxygen and pure nitrogen mixtures. A non-equilibrium factor was determined using insights from high-fidelity StS and DMS data to model the QSS dissociation phenomenon. The proposed model reduction was applied to the StS kinetics for diatom-atom interactions while a coupling of model reduction with QCT calculations was used to determine the kinetics database for diatom-diatom interactions. The resulting 1-bin 3T model was then validated against a number of test cases for 0-D isothermal and adiabatic reactor conditions. The 0-D results showed that the separation of internal energy into rotational and vibrational components accurately predicts the energy transfer process thereby highlighting the importance of modeling non-equilibrium behavior with multiple internal temperatures. Rotational and vibrational relaxation time constants for both expanding and compressing flows were compared against StS data for diatom-atom interactions with the results highlighting the ability of the proposed 1-bin 3T model to perform well regardless of initial conditions and relaxation direction. The validity of non-equilibrium factor across a range of temperature conditions was also demonstrated through the 0-D isothermal and adiabatic reactor cases. To conclude, the application of 1-bin 3T model to different test cases demonstrated that this model has the potential to be both computationally cost-efficient and highly accurate for thermochemical non-equilibrium modeling of hypersonic flows.

\begin{acknowledgements}
    The authors would like to acknowledge funding support from the Air Force Office of Scientific Research (AFOSR) under AFOSR Grant No.: FA9550-22-1-0039.
\end{acknowledgements}

\section*{Data Availability Statement}
The data that support the findings of this study are available from the corresponding author upon reasonable request.

\section*{Author Declarations}
The authors have no conflicts to disclose.

\section*{Appendix}
\appendix

\section{Statistical uncertainties in QCT calculations involving diatom-diatom interactions}

The statistical uncertainties associated with the rate coefficients obtained from quasi-classical trajectory (QCT) calculations are examined in this appendix. The QCT calculations were performed at 25 discrete collision energies ranging from 0.00367 to 1.10248 Hartree. A nonuniform collision-energy grid was employed, with finer spacing at lower collision energies and progressively coarser spacing at higher energies. A minimum of $5\times10^{5}$ trajectories were simulated at each collision energy for every combination of rotational and vibrational temperatures considered, resulting in a minimum total of 12.5 million trajectories per temperature condition. Additional trajectories were simulated at selected collision energies to improve statistical convergence. The collision-energy-dependent coefficients were subsequently integrated over the Boltzmann distribution of relative collision energies to obtain the corresponding macroscopic coefficients \cite{Kim_Hydrogen}. The statistical uncertainties were estimated by propagating the sampling variances through the Boltzmann integration. The error bars in the following figures represent one standard deviation of the estimated coefficients. Only the results for $\mathrm{N_{2}+N_{2}}$ interactions are presented as the $\mathrm{O_{2}+O_{2}}$ chemical system exhibited substantially smaller uncertainties.

\begin{figure}[h]
    \centering
    \includegraphics[width=1.1\linewidth]{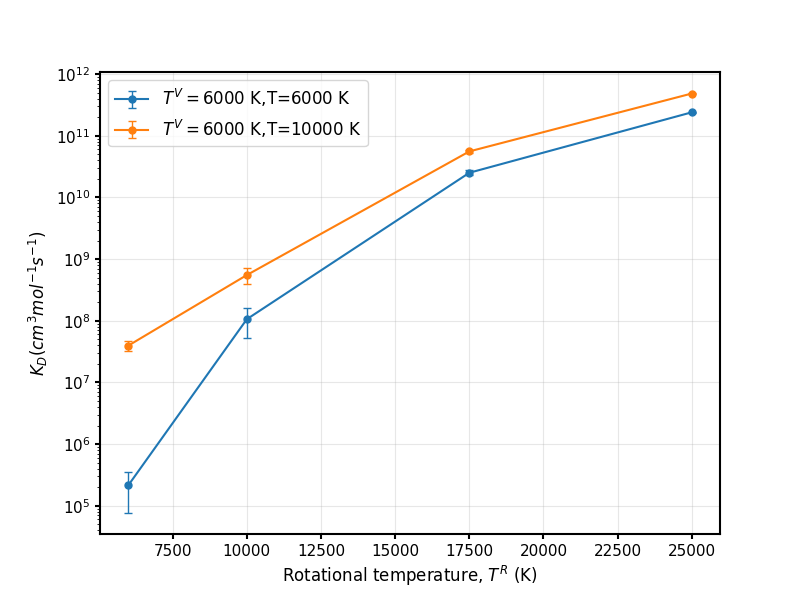}
    \caption{\textbf{Dissociation rate coefficients as functions of rotational temperature for selected translational and vibrational temperatures for $\mathrm{N_{2}+N_{2}}$ interactions.}}
    \label{fig:Diss_error}
\end{figure}

\begin{figure}[h]
    \centering
    \includegraphics[width=1\linewidth]{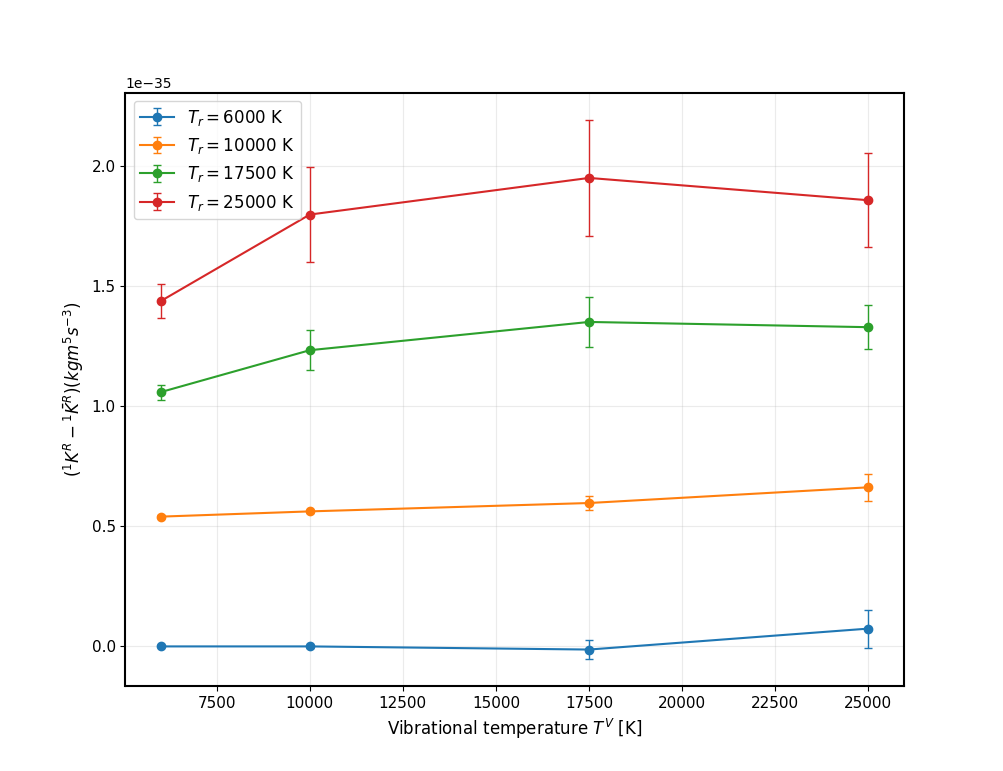}
    \caption{\textbf{$(\mathbf{{}^{1}K^{R}-{}^{1}\Bar{K}^{R}})$ values at T=6000 K for $\mathrm{N_{2}+N_{2}}$ interactions.}}
    \label{fig:Rot_SD}
\end{figure}

\begin{figure}[h]
    \centering
    \includegraphics[width=1\linewidth]{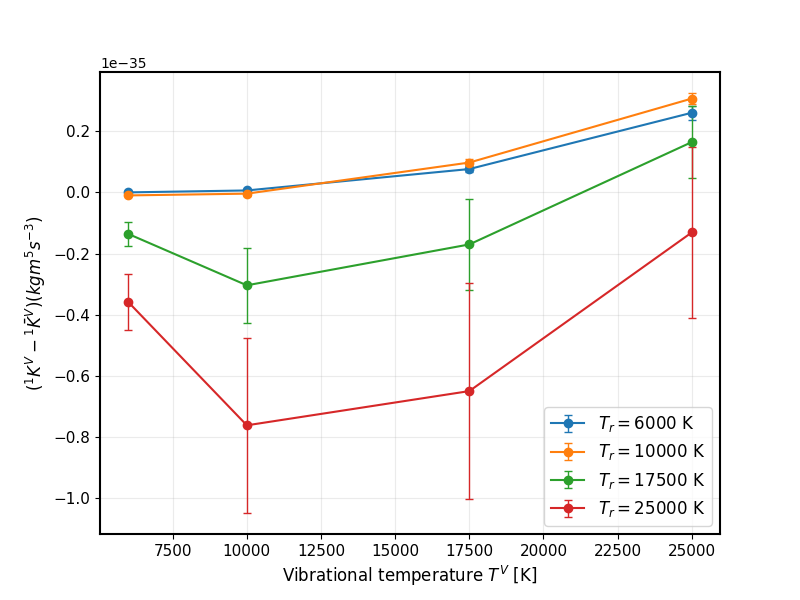}
    \caption{\textbf{$(\mathbf{{}^{1}K^{V}-{}^{1}\Bar{K}^{V}})$ values at T=6000 K for $\mathrm{N_{2}+N_{2}}$ interactions.}}
    \label{fig:Vib_SD}
\end{figure}

Figures \ref{fig:Diss_error}-\ref{fig:Vib_SD} present the statistical uncertainties associated with the dissociation, rotational energy-transfer, and vibrational energy-transfer coefficients, respectively, for selected temperature combinations for $\mathrm{N_{2}+N_{2}}$ interactions. The figures focus on conditions exhibiting appreciable statistical uncertainties within the temperature grid considered.

As shown in Fig. \ref{fig:Diss_error}, the statistical uncertainties associated with the dissociation rate coefficients are generally small, although appreciable uncertainties are observed at lower rotational temperatures, particularly for the lowest translational temperature considered (6000 K).It is worth noting that these rate coefficients do not include the non-equilibrium factor.

The statistical uncertainties associated with the rotational and vibrational energy-transfer coefficients are shown in Figs. \ref{fig:Rot_SD} and \ref{fig:Vib_SD}, respectively. The figures show that larger uncertainties are observed at elevated rotational temperatures, particularly at $T^{R}$=25000 K. The evaluation of these coefficients involves energy-weighted contributions from state-to-state transitions, including reverse processes reconstructed using detailed balance. The application of detailed balance can amplify the absolute uncertainties associated with individual transition contributions, while the combination of positive and negative energy-transfer contributions can further increase the relative uncertainty of the resulting coefficients. The largest uncertainties occur under strongly nonequilibrium conditions characterized by relatively low translational temperatures and high rotational temperatures. These temperature combinations lie outside the thermodynamic regimes encountered in the zero-dimensional reactor simulations considered in this work. Consequently, the elevated uncertainties at these conditions are not expected to substantially affect the model-validation results.

\bibliography{aipsamp}
\end{document}